\documentclass[12pt]{article}
\usepackage{geometry}             
\usepackage{graphicx}
\usepackage{amssymb}
\usepackage{amsmath}
\usepackage{epstopdf}
\usepackage{subfigure}
\usepackage{cite}

\usepackage[hyperindex=true,
          pdfstartview=FitH,
          bookmarksnumbered=true,
          bookmarksopen=true,
          citecolor=blue,
          linkcolor=blue,
          colorlinks=true,
          pdfborder=001,
          unicode]{hyperref}

\begin{document}

\title{Black holes of dimensionally continued gravity coupled to Born-Infeld electromagnetic field}
\author{Kun Meng, Da-Bao Yang\\
School of Science, Tianjin Polytechnic University, \\
Tianjin 300387, China\\
emails: \href{mailto:mengkun@tjpu.edu.cn}{\it mengkun@tjpu.edu.cn}
and \href{mailto:bobydbcn@163.com}{\it bobydbcn@163.com}}
\date{}                             
\maketitle

\begin{abstract}
In this paper, for dimensionally continued gravity coupled to Born-Infeld electromagnetic field, we construct topological black holes in diverse dimensions and construct dyonic black holes in general even dimensions.
We study thermodynamics of the black holes and obtain first laws. We study thermal phase transitions of the black holes in $T$-$S$ plane and find van der Waals-like phase transitions for even-dimensional spherical black holes, such phase transitions are not found for other types of black holes constructed in this paper .
\end{abstract}

\section{Introduction}
Since general relativity(GR) is non-renormalizable, researchers are motivated to study higher derivative gravities. It was found that higher derivative corrections to Einstein-Hilbert action can lead to a power-counting renormalizable theory\cite{Stelle1977}.
Among the modified gravities, Lovelock gravity was constructed by Lovelock with the original motivation of finding general divergence free symmetric rank-2 tensors which contain only metric and its first two derivatives\cite{Lovelock}.  Lovelock found that the desired theory of gravity consists of the dimensionally continued Euler characteristics. Because of the Lovelock tensors containing no more than 2nd order derivatives of the metric, the linearized Lovelock gravity around Minkowski spacetime is free of ghost\cite{Marugan92,Aiello04}. Lovelock gravity also emerges in string theory as effective field theory in the low energy limit, it is the higher order corrections to Einstein gravity\cite{Wiltshire86,Schwarz74}.
Since Lovelock gravity contains a lot of Lovelock coefficients which makes it difficult to extract physical information from the solution of equations of motion(EOM), Ba\~{n}ados, Teitelboim and Zanelli proposed a choice of the Lovelock coefficients, which enables one to write the solution in an explicit elegant form. The special choice of Lovelock coefficients leads to a theory of gravity which is called dimensionally continued gravity(DCG)\cite{BTZ93}. The Lovelock coefficients were properly chosen so that DCG possesses a unique AdS vacuum, and the only parameters of the theory are  gravitational constant and AdS radius\cite{Crisostomo00}. Neutral and charged black hole solutions of DCG have been found in\cite{BTZ93,Crisostomo00,Cai98}, thermodynamics of the black holes have been studied in \cite{BTZ93,Crisostomo00,Cai98,Cai06,Aiello04,MiskovicOlea}. Also, DCG black holes with scalar hair have been found in\cite{Giribet14,Giribet15}.

When we describe the dynamics of electromagnetic field, we often adopt the Maxwell theory, and usually the Maxwell theory explains electromagnetic phenomena successfully. However, it is noted that when the field is strong enough the linear Maxwell theory does not work well. In order to describe the phenomena of quantum electrodynamics, in 1936 Heisenberg and Euler proposed a nonlinear electromagnetic theory\cite{Heisenberg}. Nonlinear theories of electromagnetic field also arise in Kaluza-Klein reduction of higher-dimensional theories which include dimensionally continued Euler densities\cite{Wiltshire88,Hoissen88}. In 1930's, motivated by obtaining a finite value of the self-energy of electron Born and Infeld proposed a non-linear electrodynamics, which is know as Born-Infeld(BI) electrodynamics now\cite{BI}. In Ref.\cite{Fradkin}, Fradkin and Yeltsin showed that BI action arises naturally from string theory. The D3-brane dynamics was also noticed to be governed by BI action\cite{Tseytlin}. Recently, the special form of BI electrodynamics is used to construct new  theories, such as Eddington-inspired Born-Infeld theory\cite{EiBI} and Dirac-Born-Infeld inflation theory\cite{DBI}. Nowadays, BI theory has also been vastly used to study dark
energy, holographic superconductor and holographic entanglement entropy\cite{0307177,Jing1,Jing2}. Hoffmann first found a solution of Einstein gravity coupled to BI electromagnetic field\cite{Hoffmann}, which is devoid of essential singularity at the origin. Subsequently, many black hole solutions of gravity coupled to BI electromagnetic field with or without a cosmological constant were found \cite{Oliveira,FernandoKrug,Dey,CaiBI,Meng,Wiltshire,DehghaniHendi,Panah1508}. Thermodynamics of the BI black holes were studied in \cite{Zou10,Zou13,Mo,Panah1510,Panah1608,Panah1708}.

As mentioned above, since both Lovelock gravity and BI electromagnetic theory emerge in the low energy limit of string theory, if one considers string-generated corrections to gravity, it is natural to consider string-generated corrections to Maxwell theory simultaneously.
In this paper, we will construct black hole solutions of DCG coupled to BI electromagnetic field in diverse dimensions, since we do not restrict the dimensions of spacetime, this will help to understand some general properties of this kind of black holes.

Thermodynamics of black holes always attract a lot of attention, thermal phase transitions of black holes have been studied intensively in recent years. The initial work on thermal phase transition of black hole investigated phase transition between Schwarzchild-AdS black hole and AdS vacuum\cite{HawkingPage}, it is known as Hawking-Page phase transition. Later on,  phase transitions of black hole in  inverse temperature-horizon($1/T$-$r_{+}$) plane and temperature-entropy($T$-$S$) plane were studied\cite{Chamblin9902,Chamblin9904}, and it was found that the phase transition behavior of the black hole resembles the one of van der Waals liquid-gas system. Van der Waals-like phase transitions in $T$-$S$ plane were also observed in\cite{Mahapatra15,Mahapatra16,Zeng,Kuang}. Recently, thermal phase transition of  Reissner-N\"{o}rdstrom (RN) anti-de Sitter black hole has been studied in extended phase space by identifying the cosmological constant as pressure and the volume enclosed by horizon  as thermodynamical volume of the gravitational system \cite{Mann1205}. The cosmological constant was viewed as pressure originally  motivated by making first law be consistent with Smarr relation \cite{0904.2765}. In extended phase space van der Waals-like phase transition of RN-AdS black hole was found too. Studies of black hole thermal phase transitions in extend phase space have been generalized to various black holes\cite{Mann1612,Zou1612,Zou1702,Mo}. We will study thermal phase transitions of the black holes constructed in this paper in $T$-$S$ plane, since it was argued that the thermal phase transition behaviors in $T$-$S$ plane and in $P$-$V$ plane are dual to each other\cite{Spallucci}, also it is technically easier for us to study phase transitions in $T$-$S$ plane than in $P$-$V$ plane\cite{Kuang}.

The paper is organized as follows. In section \ref{section2}, we construct topological black hole solutions of DCG coupled to BI electromagnetic field and study thermodynamics of the black holes. In section \ref{section3}, we construct dyonic planar black hole solutions of DCG coupled to BI electromagnetic field in general even dimensions and study the thermodynamics. We conclude our results in the last section.

\section{Topological black holes\label{section2}}
\subsection{Local solution}
The action of Lovelock gravity coupled to BI electromagnetic field is given by:
\begin{align}
I=\frac{1}{16\pi G}\int \mathrm{d}^dx \sqrt{-g}\left[\sum_{p=0}^{n-1}\frac{\alpha_p}{2^p}\delta^{\mu_1\cdots\mu_{2p}}_{\nu_1\cdots\nu_{2p}}R^{\nu_1\nu_2}_{\mu_1\mu_2}\cdots R^{\nu_{2p-1}\nu_{2p}}_{\mu_{2p-1}\mu_{2p}}+L(F)\right]\label{action},
\end{align}
where $G$ is the gravitational constant, $\delta^{\mu_1\cdots\mu_{2p}}_{\nu_1\cdots\nu_{2p}}$ is the generalized Kronecker delta of order $2p$, and
\begin{align}
L(F)=4\beta^2\left(1-\sqrt{1+\frac{F_{\mu\nu}F^{\mu\nu}}{2\beta^2}}\right).\label{LF1}
\end{align}
The coefficients $\alpha_p$ in (\ref{action}) are arbitrary constants, in the special case of DCG, the $\alpha_p$'s are chosen as \cite{Kuang,BTZ93}
\begin{align}
\alpha_p=\left(
                 \begin{array}{c}
                   n-1 \\
                   p \\
                 \end{array}
               \right)\frac{(d-2p-1)!}{(d-2)!l^{2(n-p-1)}}.\label{alphap}
\end{align}
Note that, DCG  becomes Born-Infeld gravity in even dimensions and Chern-Simons gravity in odd dimensions\cite{BTZ93,Crisostomo00}.

Taking variation of the metric we obtains the EOM
\begin{align}
\sum_{p=0}^{n-1}\frac{\alpha_p}{2^{p+1}}\delta^{\nu\lambda_1\cdots\lambda_{2p}}_{\mu\rho_1\cdots\rho_{2p}}R^{\rho_1\rho_2}_{\lambda_1\lambda_2}\cdots R^{\rho_{2p-1}\rho_{2p}}_{\lambda_{2p-1}\lambda_{2p}}=T^\nu_\mu, \label{EOMG}
\end{align}
with the energy momentum tensor
\begin{align}
T^\nu_\mu=\frac{1}{2}\delta^\nu_\mu L(F)+\frac{2F_{\mu\lambda}F^{\nu\lambda}}{\sqrt{1+\frac{F_{\rho\sigma}F^{\rho\sigma}}{2\beta^2}}}.\label{energymomentum}
\end{align}
The EOM of electromagnetic field reads
\begin{align}
\partial_\mu\left(\frac{\sqrt{-g}F^{\mu\nu}}{\sqrt{1+\frac{F^{\rho\sigma}F_{\rho\sigma}}{2\beta^2}}}\right)=0.\label{EOMEM}
\end{align}
We take the static metric ansatz
\begin{align}
ds^2=-f(r)dt^2+\frac{dr^2}{f(r)}+r^2\left(h_{ij}dx^idx^j\right),\label{ansatz}
\end{align}
where $h_{ij}dx^idx^j$ is the metric of $(d-2)$-dimensional hypersurface with constant curvature.
Under the electrostatic potential assumption, all other components of the strength tensor $F^{\mu\nu}$ vanish except $F_{rt}$. Solving (\ref{EOMEM}) we have
\begin{align}
F_{rt}=\frac{Q}{\sqrt{r^{2d-4}+\frac{Q^2}{\beta^2}}}.\label{strength}
\end{align}
If (\ref{energymomentum}), (\ref{ansatz}) and (\ref{strength}) are substituted into (\ref{EOMG}), the EOM reduces to
\begin{align}
\left[\sum_{p=0}^{n-1}\frac{(d-2)!\alpha_p}{2(d-2p-1)!}r^{d-1}\left(\frac{k-f(r)}{r^2}\right)^p\right]'=32 \pi G \beta^2 r^{d-2} \left(\sqrt{1+\frac{Q^2r^{4-2d}}{\beta^2}}-1\right).
\end{align}
Where $'$ denotes derivative with respect to $r$,  $k=1,0,-1$ corresponds respectively to the co-dimension-2 hypersurface with spherical, planar or hyperbolic topology. If $\alpha_p$ takes the value given in Eq.(\ref{alphap}), the EOM is simplified to be an elegant form
\begin{align}
\left[r^{d-1}\left(\frac{1}{l^2}+\frac{k-f(r)}{r^2}\right)^{n-1}\right]'=64 \pi G \beta^2 r^{d-2} \left(\sqrt{1+\frac{Q^2r^{4-2d}}{\beta^2}}-1\right).
\end{align}
Finally, we obtain the black hole solution
\begin{align}
f(r)=k+\frac{r^2}{l^2}-r^2\left(\frac{16\pi GM}{\Sigma_{d-2}r^{d-1}}+\frac{\delta_{d,2n-1}}{r^{d-1}}-\Xi\right)^{\frac{1}{n-1}},\label{solution}
\end{align}
with
\begin{align}
\Xi=\frac{64\pi G \beta^2}{d-1}\bigg(1-\frac{\sqrt{Q^2r^4+\beta^2r^{2d}}}{\beta r^d}
+\frac{(d-2)Q^2}{(d-3)\beta^2r^{2d-4}}\cdot\;_2F_1\left[\frac{1}{2},\frac{d-3}{2(d-2)},\frac{7-3d}{4-2d},\frac{-Q^2r^{4-2d}}{\beta^2}\right]\bigg).
\end{align}
Where $M$ is an integration constant which represents mass of the black hole, $\Sigma_{d-2}$ is the volume of $(d-2)$-dimensional hypersurface. The addition of $\delta_{d,2n-1}$ in (\ref{solution}) is because the black hole horizon is expected to shrinks to a point for $M\rightarrow0$\cite{Crisostomo00}.

It's easy to check that, the black hole (\ref{solution}) degenerates to charged black hole of DCG \cite{BTZ93,Cai98,Kuang} in the limit $\beta\rightarrow\infty$, and it degenerates to neutral black hole of DCG  when $Q\rightarrow0$.

\begin{figure}[h]
\begin{center}
\includegraphics[width=.32\textwidth]{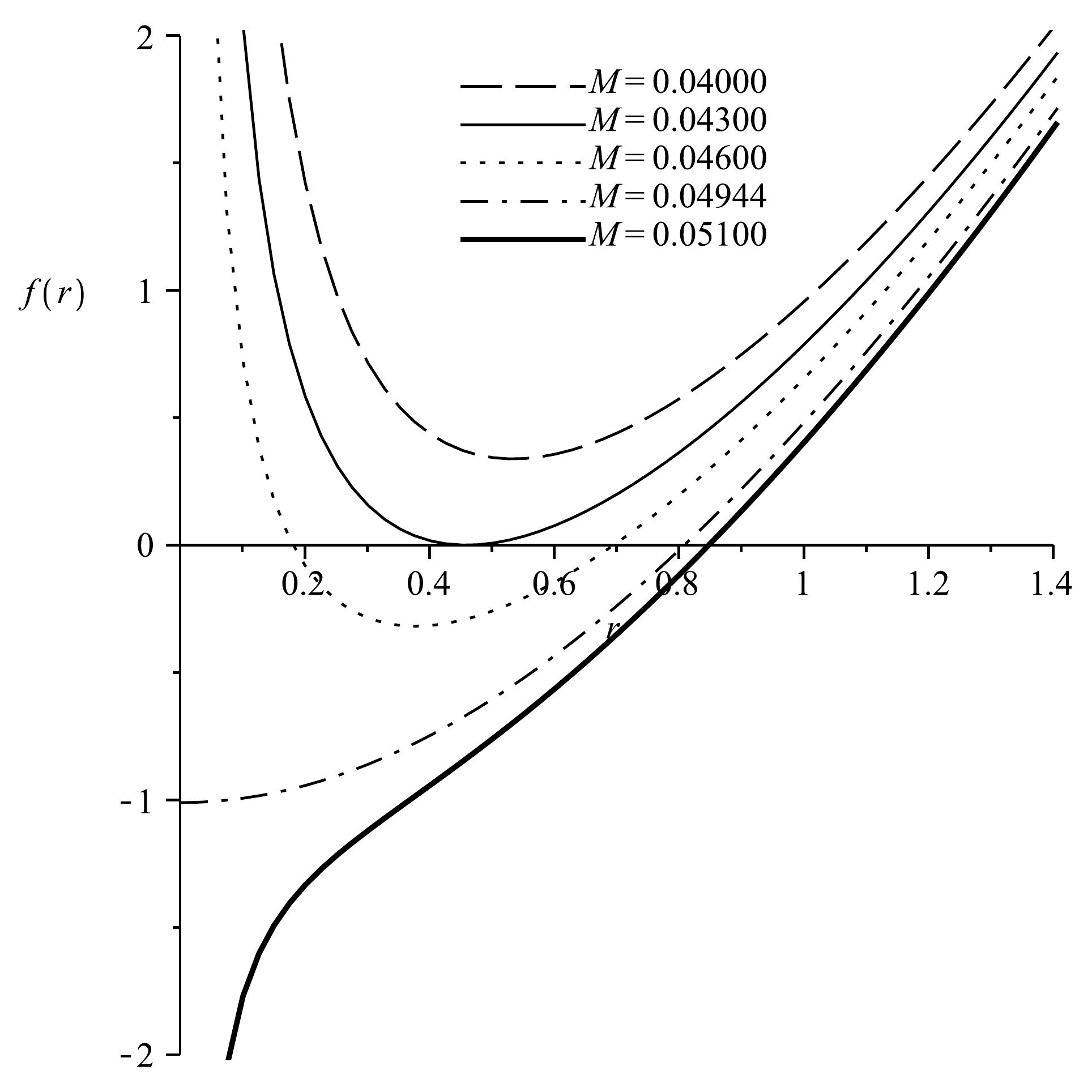}
\includegraphics[width=.32\textwidth]{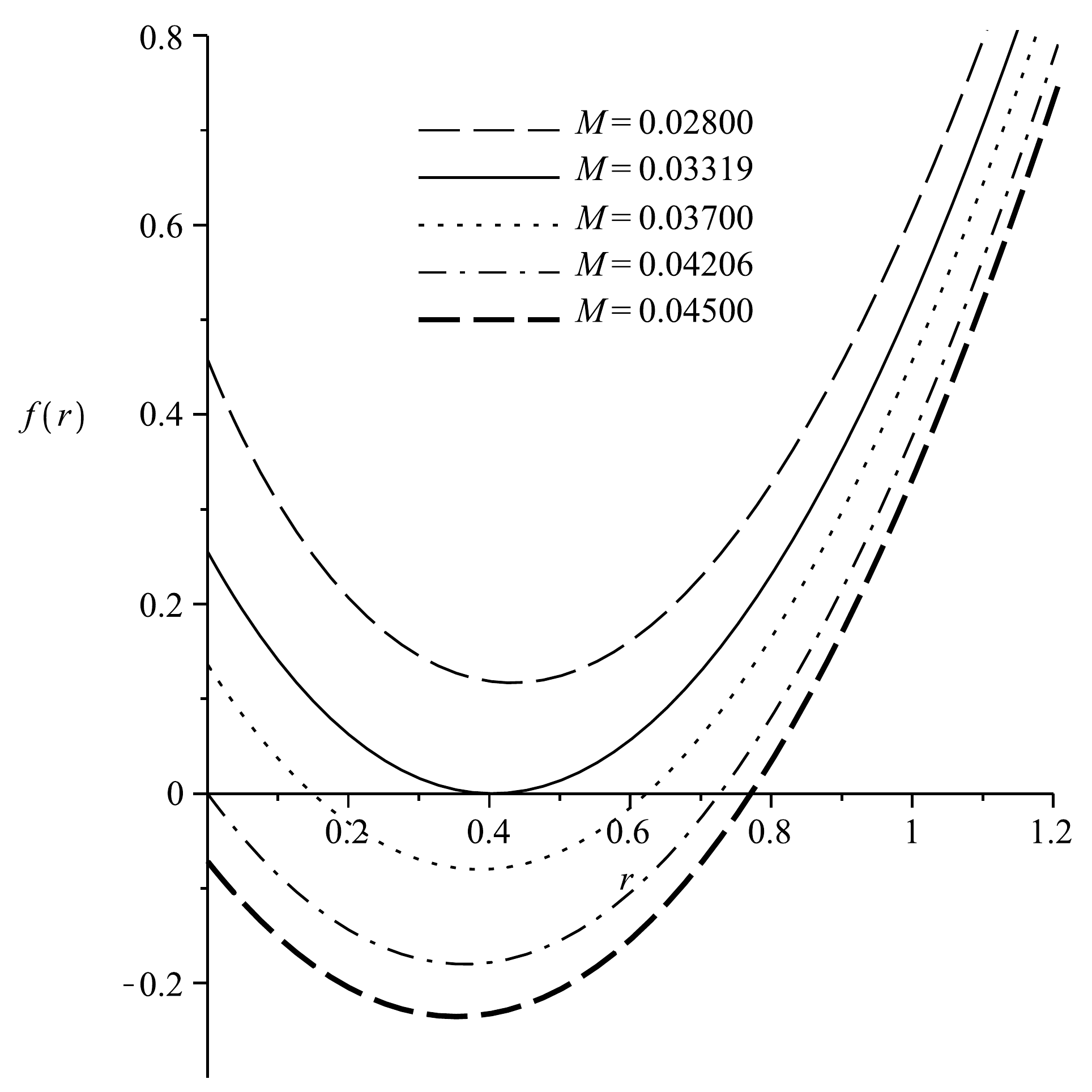}
\includegraphics[width=.32\textwidth]{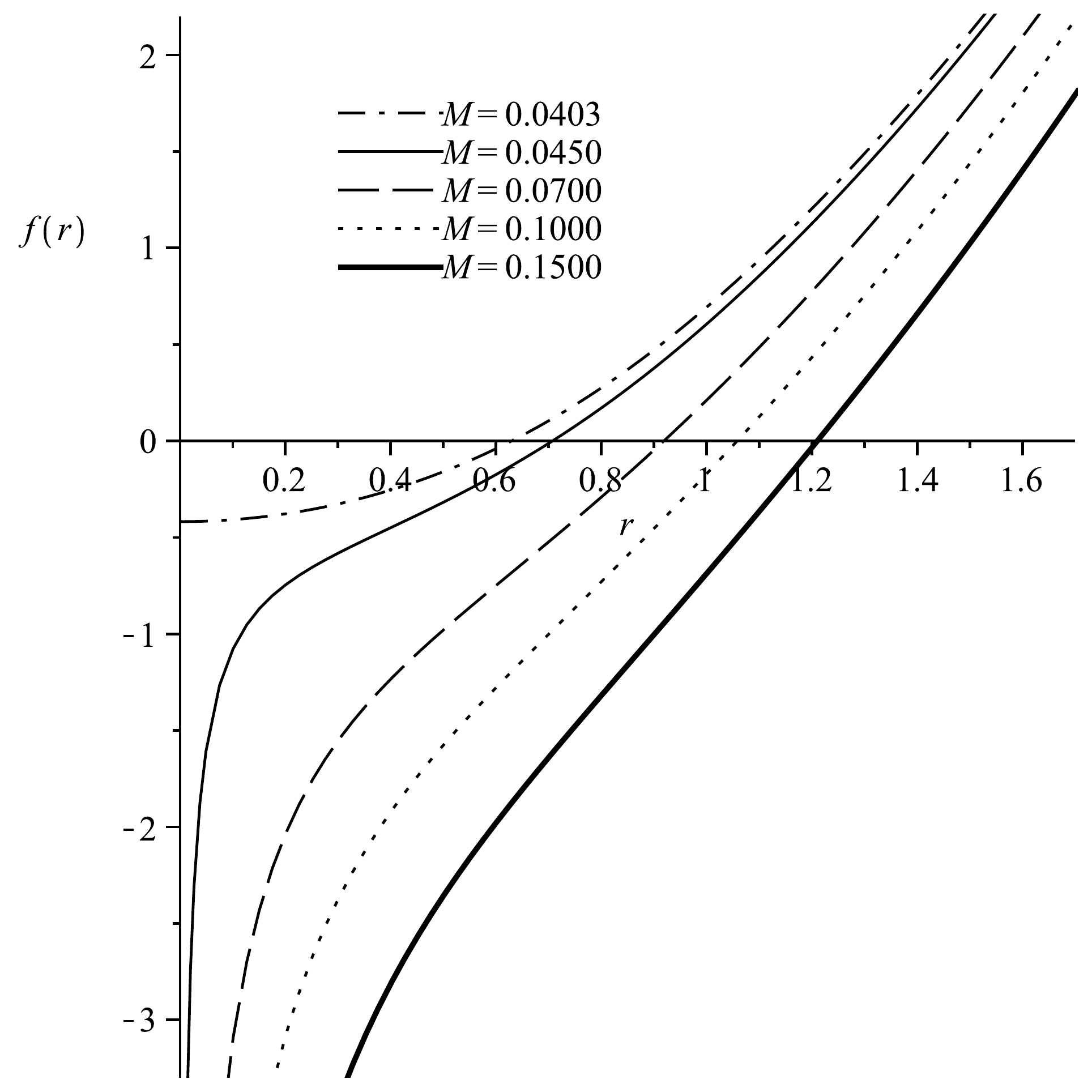}
\end{center}
\begin{center}
\includegraphics[width=.32\textwidth]{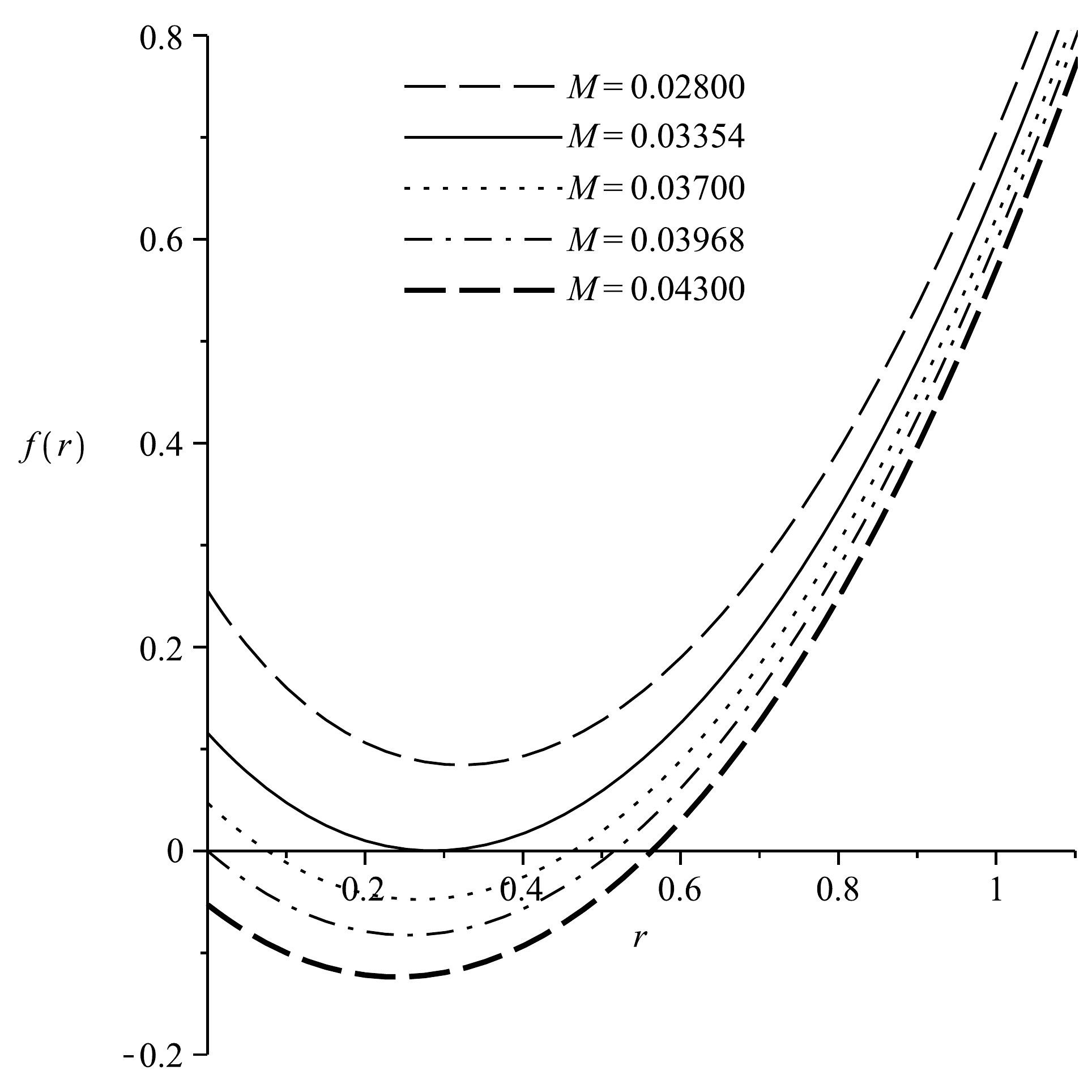}
\includegraphics[width=.32\textwidth]{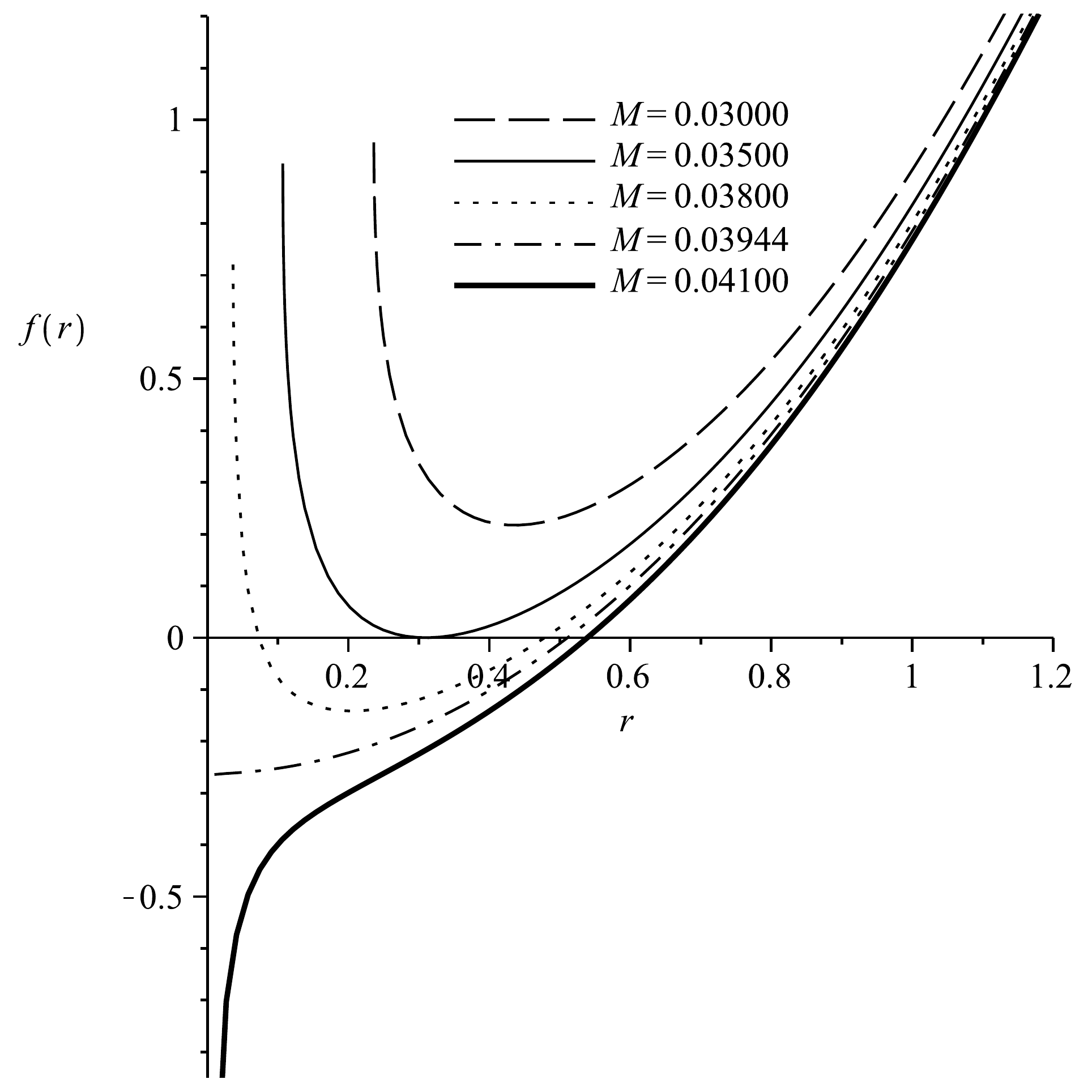}
\includegraphics[width=.32\textwidth]{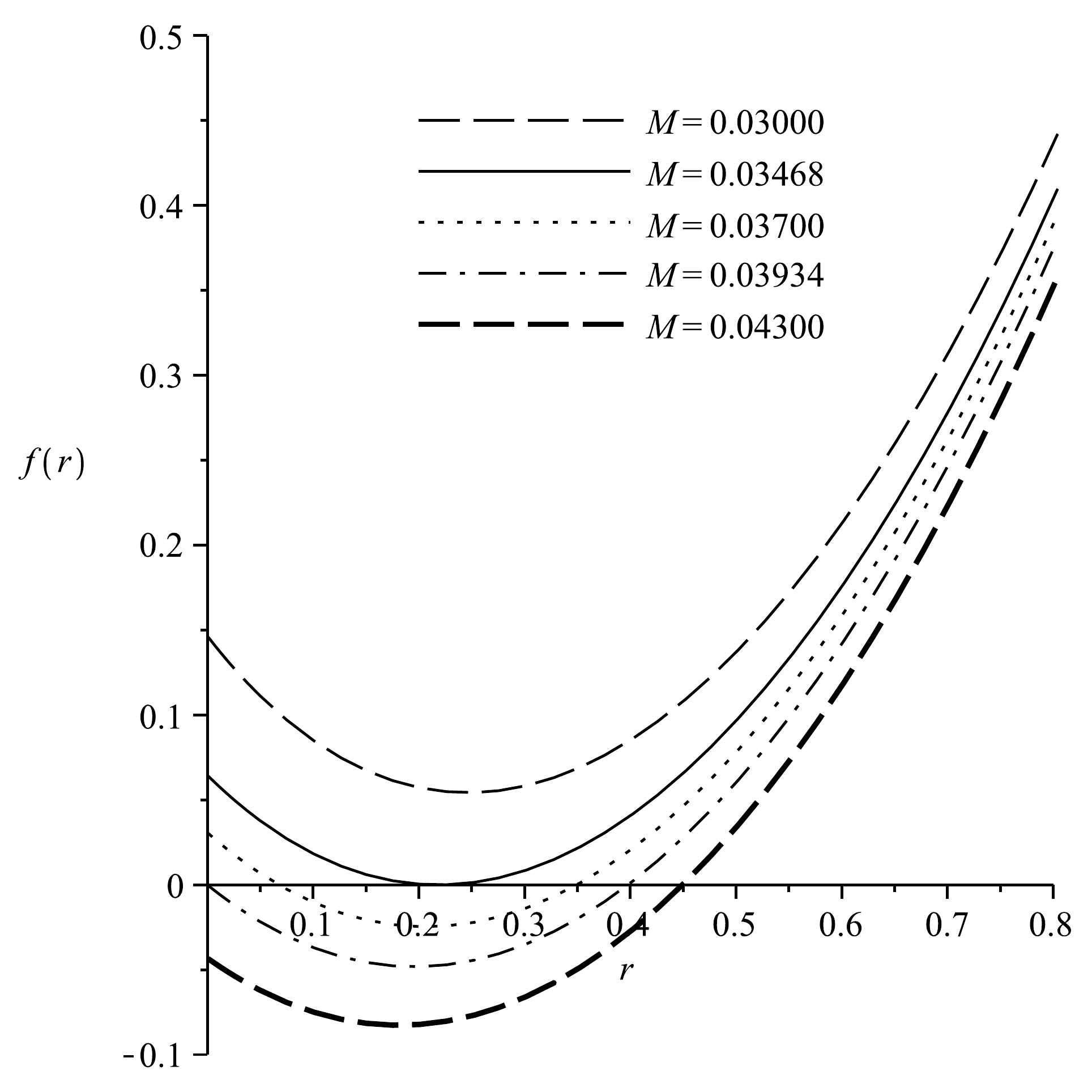}
\end{center}
\caption{$f(r)$ versus $r$. The upper three plots from left to right correspond to $d=4,5,6$ respectively. The lower three plots from left to right correspond to $d=7,8,9$ respectively. The parameters are fixed as $G=1,l=1,Q=0.1,\beta=0.1$. The dash-dotted line on every plot corresponds to $M=M_0$.}
\label{fig1}
\end{figure}

Now let's study the behaviors of $f(r)$. We take $k=1$ in the following, $k=0,-1$ can be discussed similarly.  In the limit $r\rightarrow0$, $f(r)$ is expanded  as
\begin{align}
f(r)&=1+\frac{r^2}{l^2}-\tilde{A}^{\frac{2}{d-2}}\left(\frac{1}{r^{\frac{2}{d-2}}}+\frac{2\tilde{B}}{d-2}r^{\frac{d-4}{d-2}}-\frac{(d-4)\tilde{B}^2}{(d-2)^2}r^{\frac{2(d-3)}{d-2}}
+\ldots\right),  \;\;d=2n,\label{fr0even}\\
f(r)&=1+\frac{r^2}{l^2}-\tilde{A}^{\frac{2}{d-1}}\left(1+\frac{2\tilde{B}r}{d-1}-\frac{(d-3)\tilde{B}^2r^2}{(d-1)^2}-\frac{128\pi G\beta^2}{(d-1)^2\tilde{A}}r^{d-1}+\ldots\right), \;\;d=2n-1,\label{fr0odd}
\end{align}
where
\begin{align}
\tilde{A}&=\frac{16\pi G M}{\Sigma_{d-2}}+\delta_{d,2n-1}-\frac{64\pi G\beta^2(d-2)}{\sqrt{\pi}(d-1)(d-3)}\left(\frac{Q}{\beta}\right)^{\frac{d-1}{d-2}}\Gamma\left(\frac{7-3d}{4-2d}\right)\Gamma\left(\frac{1}{2(d-2)}\right),\\
\tilde{B}&=\frac{64\pi G Q\beta}{(d-1)\tilde{A}}\left(1+\frac{2(d-2)^2}{(d-3)}\Gamma\left(\frac{7-3d}{4-2d}\right)/\Gamma\left(\frac{d-3}{2(d-2)}\right)\right).
\end{align}
From the expansions above it can be seen that, for $d=4h$ ($h=1,2,3,\ldots$), if $\tilde{A}<0$, $f(r)\rightarrow +\infty$ when $r\rightarrow0$, there may be two horizons, one horizon or no horizon (bare singularity). If $\tilde{A}>0$ , $f(r)\rightarrow-\infty$ when $r\rightarrow0$, there is a single horizon. The case $\tilde{A}=0$ will be discussed separately in the following. For $d=4h+2$, since $\tilde{A}$ must be positive in order to keep $\tilde{A}^{\frac{2}{d-2}}$ being real, we have $f(r)\rightarrow-\infty$ when $r\rightarrow0$, there is one horizon. For $d=4h+1$, if $\tilde{A}^{\frac{2}{d-2}}<1$, $f(r)$ is finite and positive at $r=0$, there may be two horizons, one horizon or no horizon. If  $\tilde{A}^{\frac{2}{d-2}}\geq1$, $f(r)$ is finite and nonpositive at $r=0$, there is  one horizon.
For $d=4h+3$, if $\tilde{A}<0$ or $0<\tilde{A}^{\frac{2}{d-1}}<1$, $f(r)$ is finite and positive at $r=0$, there may be two horizons, one horizon or no horizon. If $\tilde{A}^{\frac{2}{d-1}}\geq1$, $f(r)$ is finite and nonpositive at $r=0$, there is one horizon.
If $\tilde{A}=0$, the value of $f(r)$ at $r=0$  is not easy to see explicitly from the expansion since divergence appears ($\tilde{A}$ appears in the denominator of $\tilde{B}$), however, it can be checked that $f(r)$ is finite at $r=0$ for even-dimensional spacetime and $f(r)=1$ at $r=0$ for odd-dimensional spacetime. To see the above discuss more explicitly, we present the behaviors of $f(r)$ in Fig.\ref{fig1}.

\begin{figure}[h]
\begin{center}
\includegraphics[width=.32\textwidth]{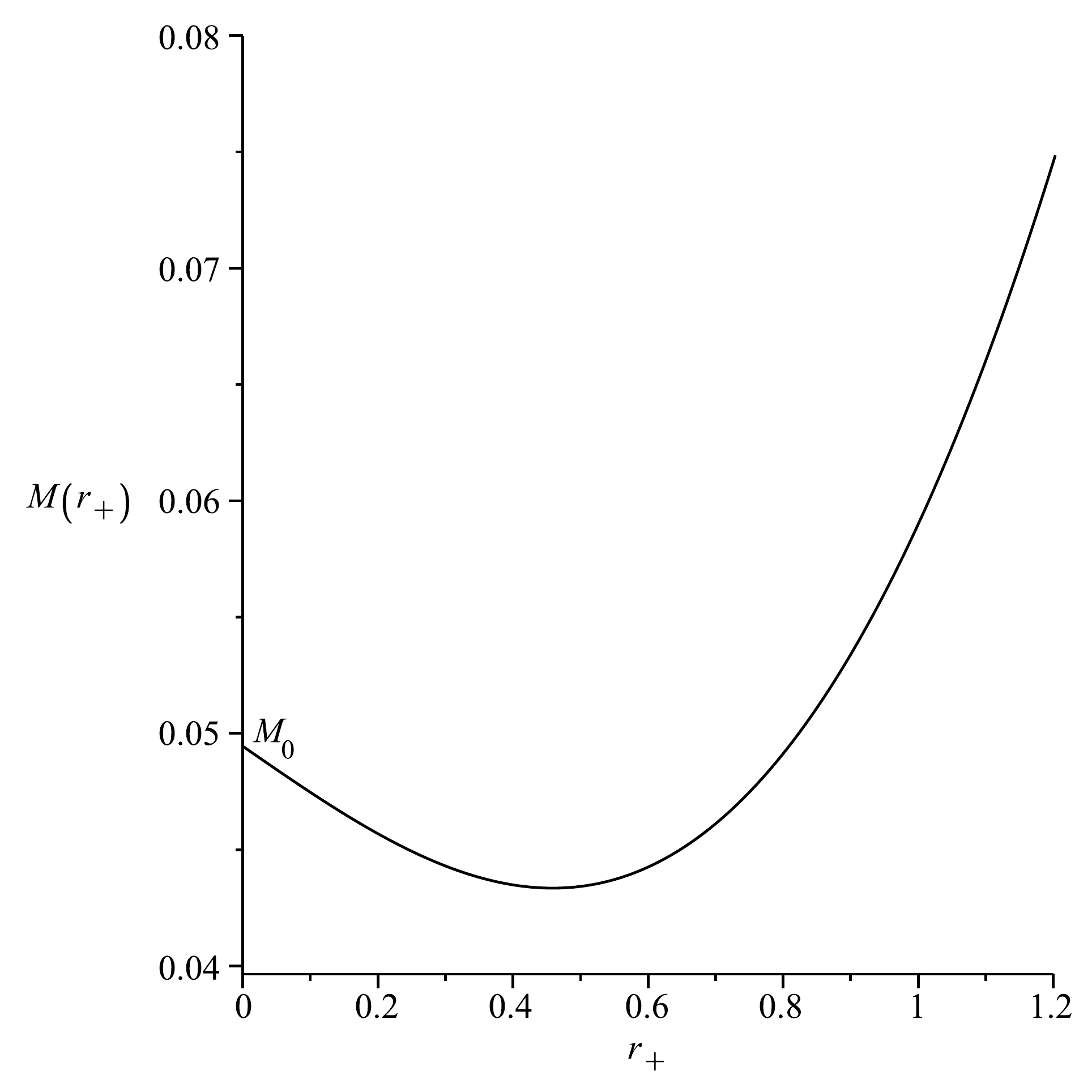}
\includegraphics[width=.32\textwidth]{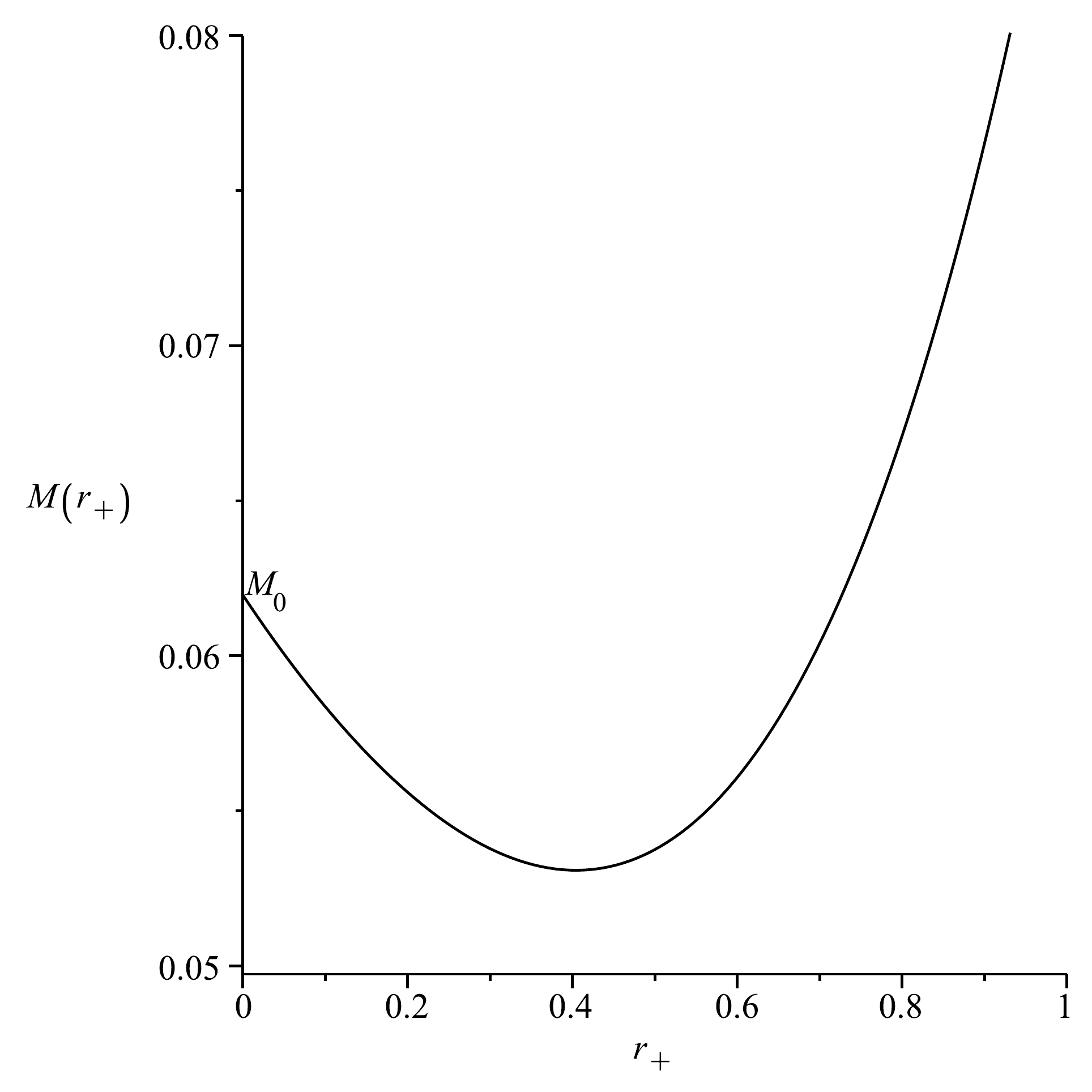}
\includegraphics[width=.32\textwidth]{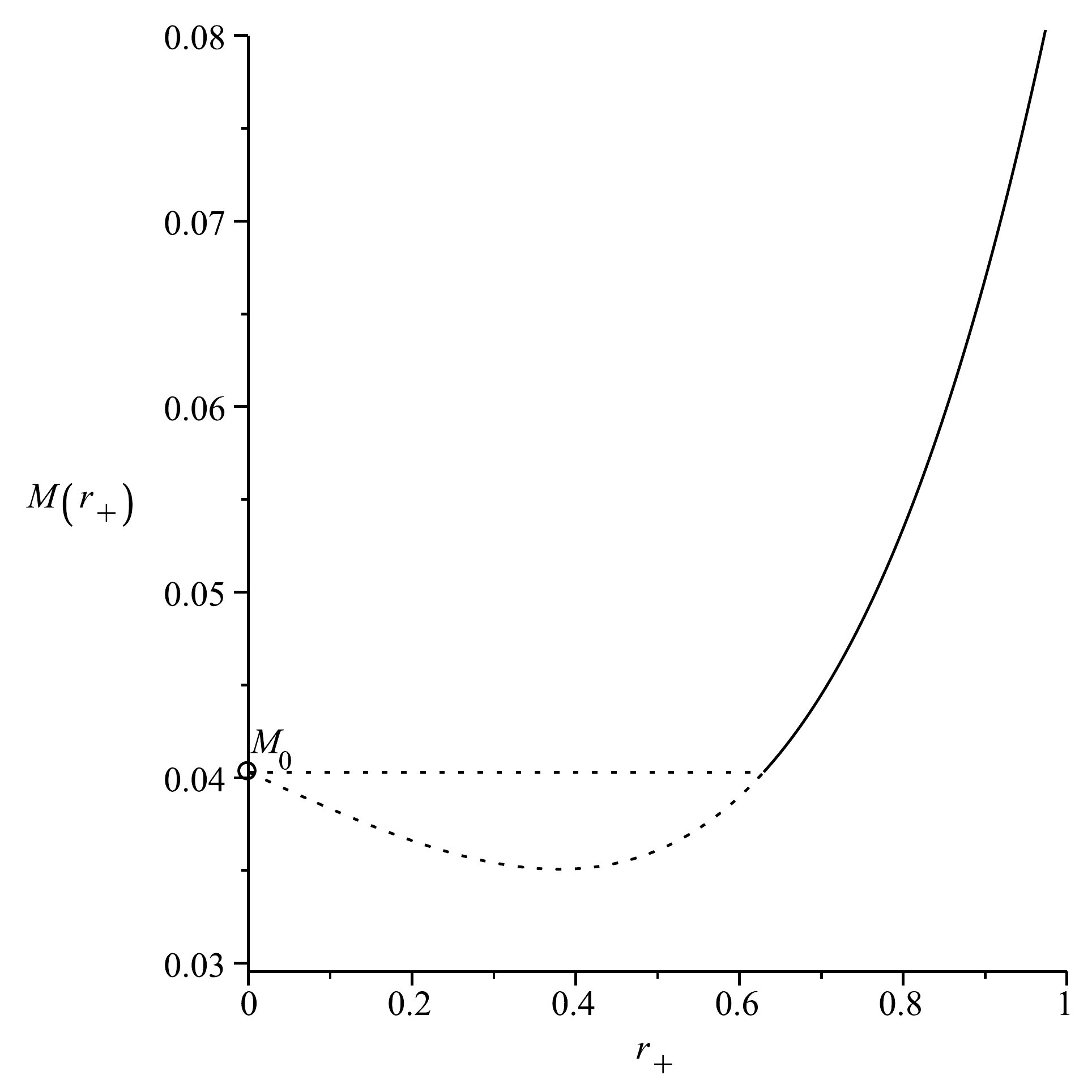}
\end{center}
\begin{center}
\includegraphics[width=.32\textwidth]{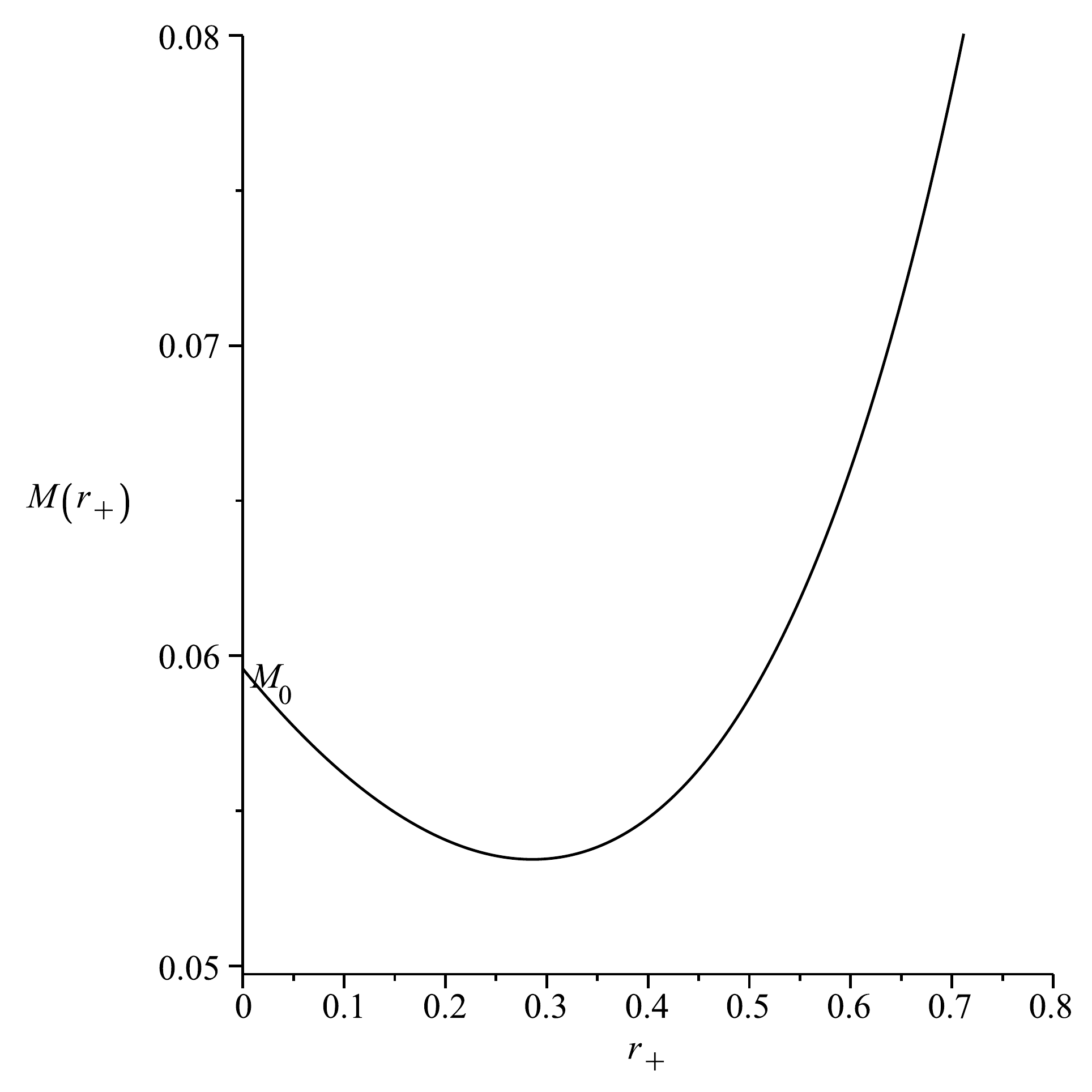}
\includegraphics[width=.32\textwidth]{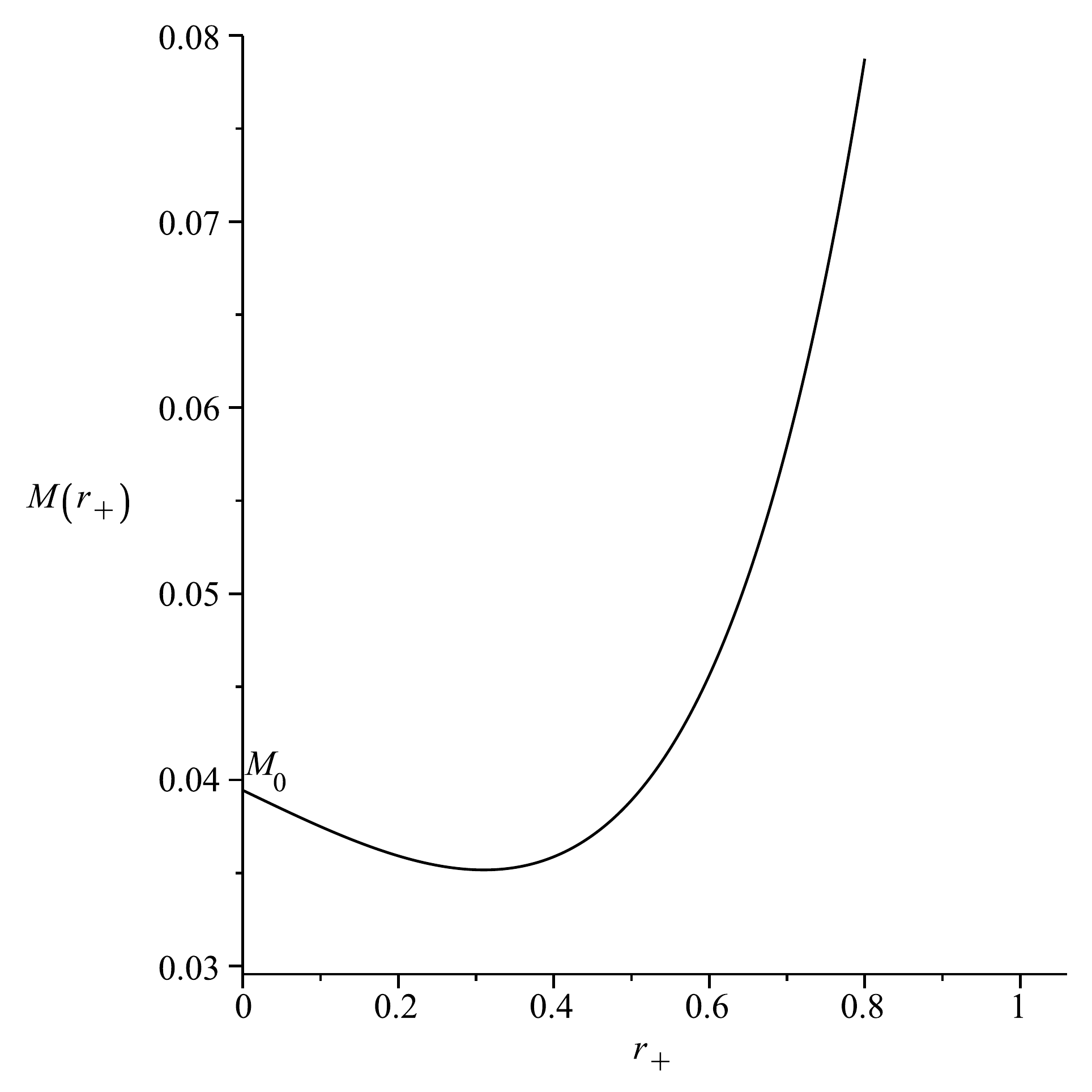}
\includegraphics[width=.32\textwidth]{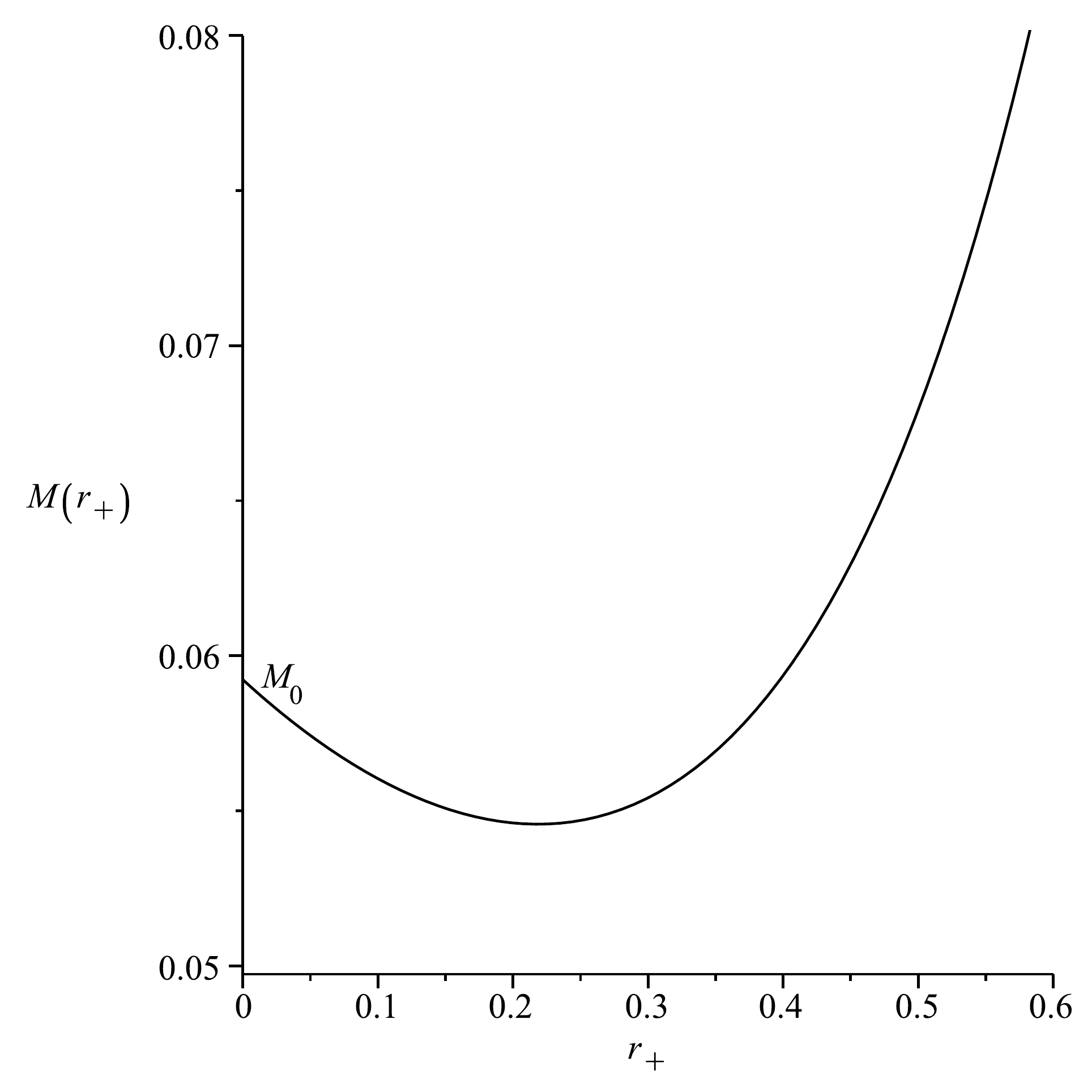}
\end{center}
\caption{$M(r_{+})$ versus $r_{+}$. These six plots are in one-to-one correspondence with the six plots in Fig.\ref{fig1}. For $d=6$, we only present the $M\geq M_0$ region (the solid line). For every other spacetime with $d\neq6$, there exists a minimum $M_{ext}$, and $M=M_0$ at $r_{+}=0$.}
\label{fig2}
\end{figure}

To understand the horizons more deeply, we plot $M$ as a function of the horizon radius $r_{+}$ in Fig.\ref{fig2}, the form of $M(r_{+})$ is given in (\ref{mass}) in the following.
As we can see, for a spacetime with $d\neq6$, there exists a minimum value $M_{ext}$ of $M$, this corresponds to an extremal black hole. When $r_{+}=0$, $M$ takes a finite positive value $M_0$. If $M_{ext}<M<M_0$, there are two horizons, if $M\geq M_0$, there is a single horizon, if $M<M_{ext}$, there is no horizon. For $d=6$ (actually $d=4h+2$), since $f(r)$ is imaginary when $M<M_0$, we only present the $M\geq M_0$ region. In the $M\geq M_0$ region, $M(r_{+})$ is a monotonic function of $r_{+}$, therefore, one given mass corresponds to one black hole horizon. Note that for even-dimensional spacetime $M=M_0$ corresponds to $\tilde{A}=0$, while for odd-dimensional spacetime $M=M_0$ corresponds to $\tilde{A}=1$.

\subsection{Thermodynamics\label{section3}}
In this section, we study  thermodynamics of the black holes constructed above. First we give the thermodynamic quantities and check the first law of thermodynamics, then we study  thermal phase transition behaviors of the black holes in $T$-$S$ plane.

\subsubsection{First law of thermodynamics}
In term of the horizon radius, mass of the black hole is given by
\begin{align}
M=&\frac{\Sigma_{d-2}}{16 \pi G  r_{+}^{d+1}(d-1)(d-3)(k l^2+r_{+}^2)}\bigg\{(d-3)r_{+}^d\bigg[64\pi G \beta r_{+}^2\left(r_{+}^d\beta-\sqrt{Q^2r_{+}^4+r_{+}^{2d}\beta^2}\right)\nonumber\\
&+(d-1)l^2r_{+}^{d+2}\left(\frac{1}{l^2}+\frac{k}{r_{+}^2}\right)^n +64k\pi G l^2\beta^2r_{+}^d-64k\pi G\beta l^2\sqrt{Q^2r_{+}^4+r_{+}^{2d}\beta^2}\bigg]\nonumber\\
&+64(d-2)\pi G Q^2r_{+}^4(kl^2+r_{+}^2)\cdot_2F_1\left[\frac{1}{2},\frac{d-3}{2(d-2)},\frac{7-3d}{4-2d},\frac{-Q^2r_{+}^{4-2d}}{\beta^2}\right]\bigg\}-\frac{\delta_{d,2n-1}\Sigma_{d-2}}{16\pi G}
.\label{mass}
\end{align}
Temperature of the black hole $T=\frac{f^{'}(r_+)}{4\pi}$ reads
\begin{align}
T&=\frac{\left(\frac{1}{l^2}+\frac{k}{r_{+}^2}\right)^{-n}}{4l^4(n-1)\pi r_{+}^{d+3} \sqrt{Q^2r_{+}^4+r_{+}^{2d}\beta^2}}\bigg[l^2r_{+}^{d+2}\left(\frac{1}{l^2}+\frac{k}{r_{+}^2}\right)^n
\left(kl^2(d+1-2n)+(d-1)r_{+}^2\right)\nonumber\\
&\cdot\sqrt{Q^2r_{+}^4+r_{+}^{2d}\beta^2}-64\pi G \beta (kl^2+r_{+}^2)^2\left(Q^2r_{+}^4+r_{+}^{d}\beta\left(r_{+}^{d}\beta-\sqrt{Q^2r_{+}^4+r_{+}^{2d}\beta^2}\right)\right)\bigg].
\end{align}
The entropy, which is defined through Wald's method, is given by
\begin{align}
S&=-2\pi\oint d^{d-2}x\sqrt{h}Y^{\mu\nu\rho\sigma}\epsilon_{\mu\nu}\epsilon_{\rho\sigma}\nonumber\\
&=\frac{\Sigma_{d-2}r_{+}^{d-2}}{2G}\sum_{p=1}^{n-1}\frac{p}{2^p}\delta^{\mu_1\cdots\mu_{2p-2}}_{\nu_1\cdots\nu_{2p-2}}\left(\alpha_pR^{\nu_1\nu_2}_{\mu_1\mu_2}\cdots R^{\nu_{2p-4}\nu_{2p-2}}_{\mu_{2p-4}\mu_{2p-2}}\right)\nonumber\\
&=\frac{(n-1)\Sigma_{d-2}r_+^d}{4 k G(d-2n+2)}\left(\frac{k}{r_+^2}+\frac{1}{l^2}\right)^{n-1}\;_2F_1\left[1,\frac{d}{2},\frac{d-2n+4}{2},-\frac{r_+^2}{kl^2}\right],
\end{align}
where $\epsilon_{\mu\nu}$ is the normal bivector of  the $t=const$ and $r=r_{+}$ hypersurface with $\epsilon_{\mu\nu}\epsilon^{\mu\nu}=-2$, and $Y^{\mu\nu\rho\sigma}\equiv\frac{\partial \mathcal{L}}{\partial R_{\mu\nu\rho\sigma}}$. The electric charge is calculated through performing integration of the flux of electromagnetic field in (\ref{EOMEM}) on the $t=const$ and $r\rightarrow\infty$ hypersurface, yielding
\begin{align}
Q_e=4Q\Sigma_{d-2}.
\end{align}
The electric potential, measured at infinity with respect to the horizon, is defined as
\begin{align}
\Phi&=A_\mu\chi^\mu|_{r\rightarrow\infty}-A_\mu\chi^\mu|_{r=r_{+}}\nonumber\\
&=\frac{q}{(d-3)r_{+}^{d-3}}\cdot\;_2F_1\left[\frac{1}{2},\frac{d-3}{2(d-2)},\frac{7-3d}{2(2-d)},-\frac{q^2r_{+}^{4-2d}}{\beta^2}\right]
\end{align}

Having the above thermodynamical quantities in hand, one can check that the first law of thermodynamics
\begin{align}
  dM=TdS+\Phi dQ_e
\end{align}
is satisfied.

\subsubsection{Thermal phase transition}
We study phase transitions of the black holes in the $T$-$S$ plane with $Q$  kept fixed.  The critical equations are

\begin{align}
\frac{\partial T}{\partial S}\bigg|_Q=0,\;\;\;\;\;\;\;\;\;\;\;\;\;\frac{\partial^2 T}{\partial S^2}\bigg|_Q=0.\label{criticaleq}
\end{align}
Since the expression of $S$ as a function of $r_{+}$ is a little tedious, it is not easy for $S(r_{+})$ to be inverted to give $r_{+}(S)$. We take the strategy
\begin{align}
\frac{\partial T}{\partial S}\bigg|_Q=\frac{\frac{\partial T}{\partial r_{+}}}{\frac{\partial S}{\partial r_{+}}}\bigg|_Q
\end{align}
to perform the derivative.

In the following, we will discuss even-dimensional ($d=2n$) and odd-dimensional ($d=2n-1$) black holes with different topologies separately.
We first study even-dimensional black holes. For spherical black holes with $k=+1$, we solve the critical equation $\frac{\partial T}{\partial S}\big|_Q$ and get the square of electric charge $Q^2$ as a function of $r_{+}$,
\begin{align}
Q^2=-r_{+}^{4(n-1)} \beta^2+\frac{\left[(lr_{+})^{2(1-n)}(k l^2+r_{+}^2)^n \left(k l^2+(2n-1)r_{+}^2\right)-64 \pi G \left(k l^2 (2n-3)+r_{+}^2\right)\beta^2+\Delta\right]^2}{16384 \pi^2 G^2 \beta^2 \left(k l^2+(2n-3)r_{+}^2\right)^2 r_{+}^{-4(n-1)}}\label{Qc2}
\end{align}
with
\begin{align}
\Delta&=\bigg\{32768 \pi^2 G^2 \beta^4 (n-1) (k l^2+r_{+}^2)^3 \left(k l^2+(2n-3)r_{+}^2\right)\nonumber\\
&\;\;\;\;+\left[l^{2(1-n)}r_{+}^{2(1-n)}(k l^2+r_{+}^2)^n \left(k l^2+(2n-1)r_{+}^2\right)-64 \pi G \beta^2 (k l^2+r_{+}^2) \left(k l^2 (2n-3)+r_{+}^2\right)\right]^2\bigg\}^{1/2}.\label{Delta}
\end{align}
Substituting (\ref{Qc2}) and (\ref{Delta}) into $\frac{\partial^2 T}{\partial S^2}\big|_Q=0$ in (\ref{criticaleq}) one gets the final form of the critical equation. The critical equation is a little lengthy we will not present it here and it is too complicated to be solved analytically, the numeral results are listed in the table.
\begin{table}
\begin{minipage}[c]{1.0\textwidth}
\centering
\begin{tabular}{cccccc}
\hline\hline
$n$ & $\beta$ & $r_{c}$ & $Q_c$ & $T_c$ & $S_c$ \\ [0.5ex]
\hline
2 & 0.1 & 4.07046 & 0.166712 & 0.025978993 & 2.07108\\
\hline
3 & 0.1 & 3.67453 & 0.357354 & 0.017466757 & 3.60344\\
\hline
4 & 0.05 & 3.29260 & 0.712754 & 0.013907099 &4.52212\\
\hline
4 & 0.1 & 3.29264 & 0.712732 & 0.013907112 &4.52226\\
\hline
4 & 1 & 3.29266 & 0.712725 & 0.013907109 &4.52231\\
\hline
\end{tabular}
\label{table1}
\caption{The critical points for different $n$ and $\beta$.}
\end{minipage}
\end{table}

\begin{figure}[h]
\begin{center}
\includegraphics[width=.30\textwidth]{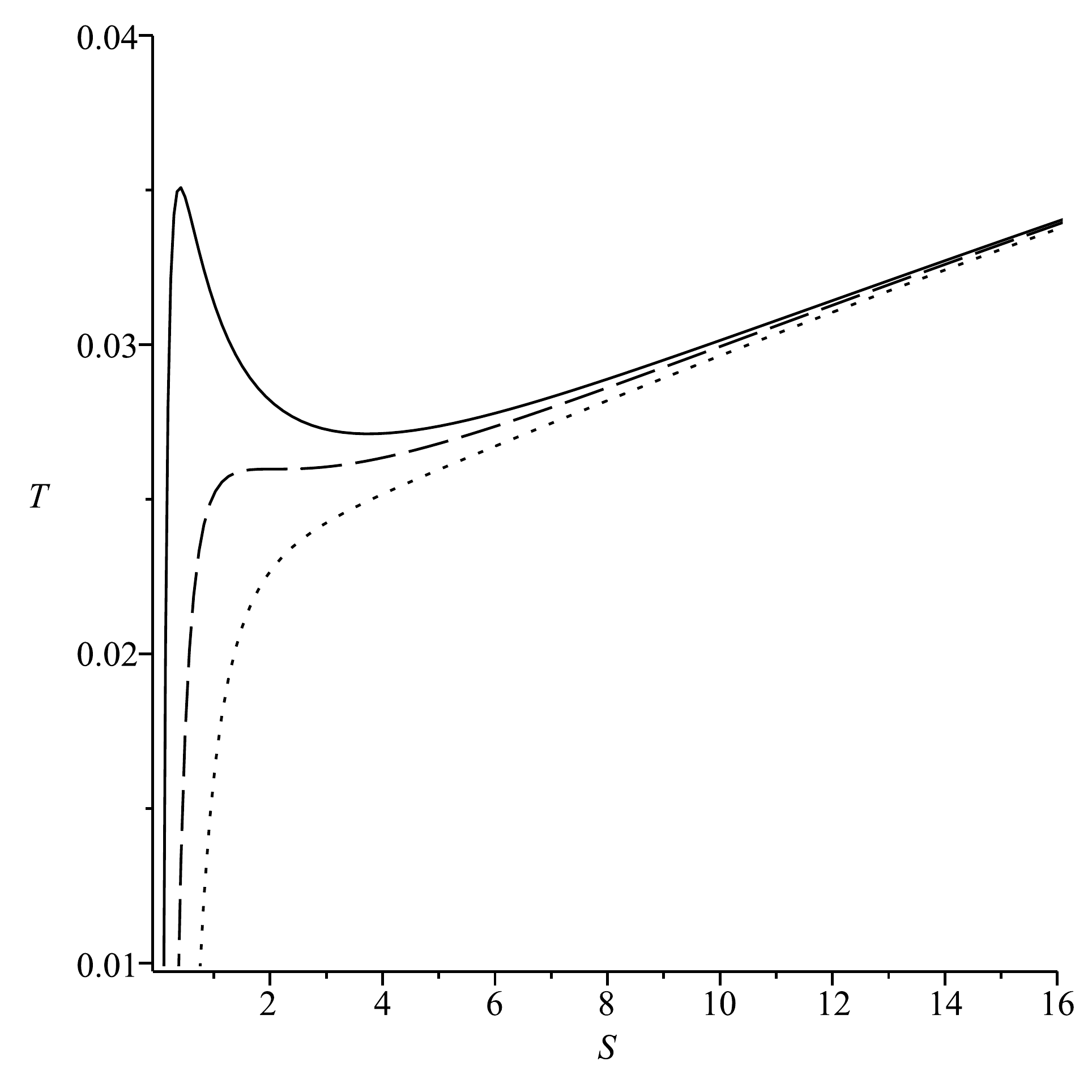}
\includegraphics[width=.30\textwidth]{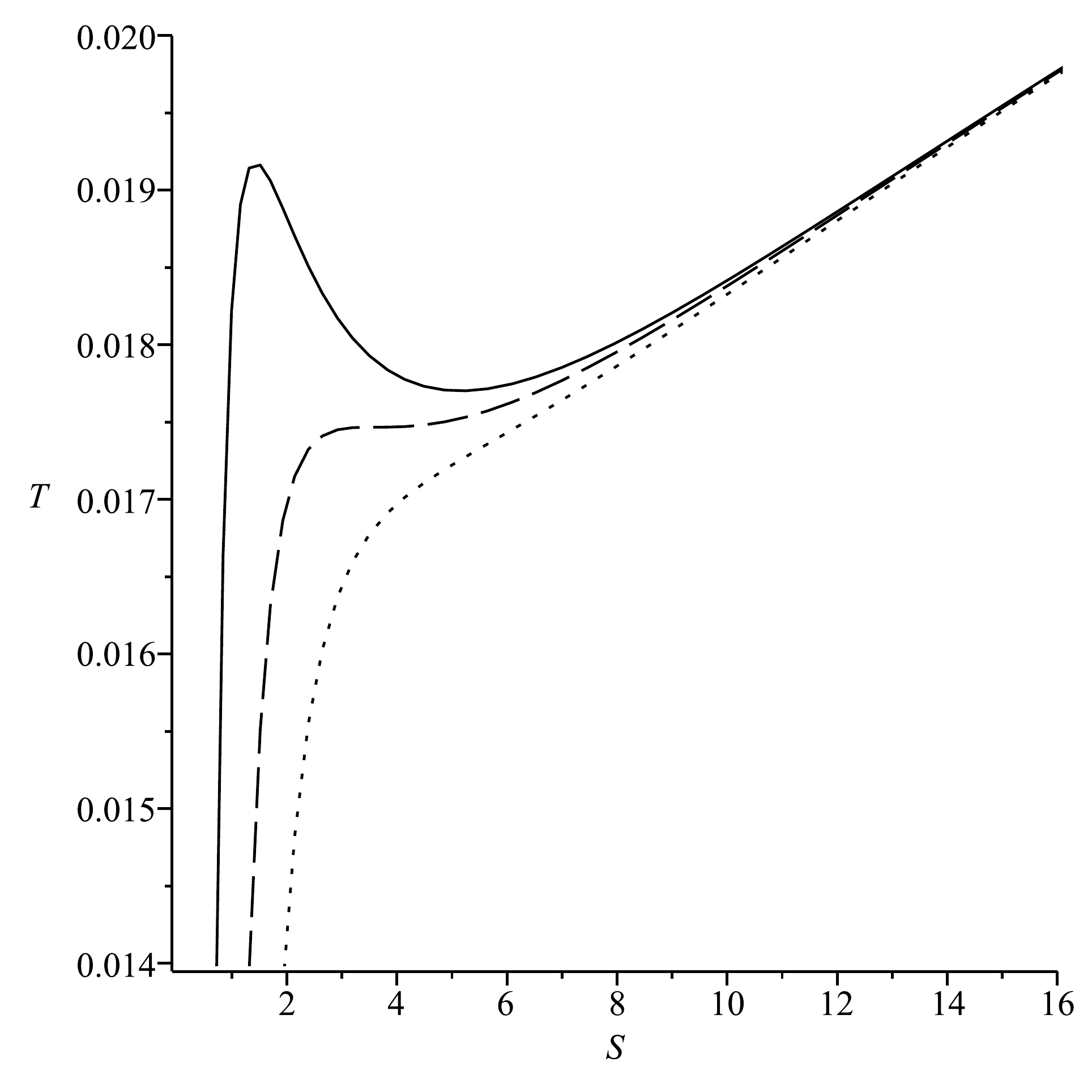}
\includegraphics[width=.30\textwidth]{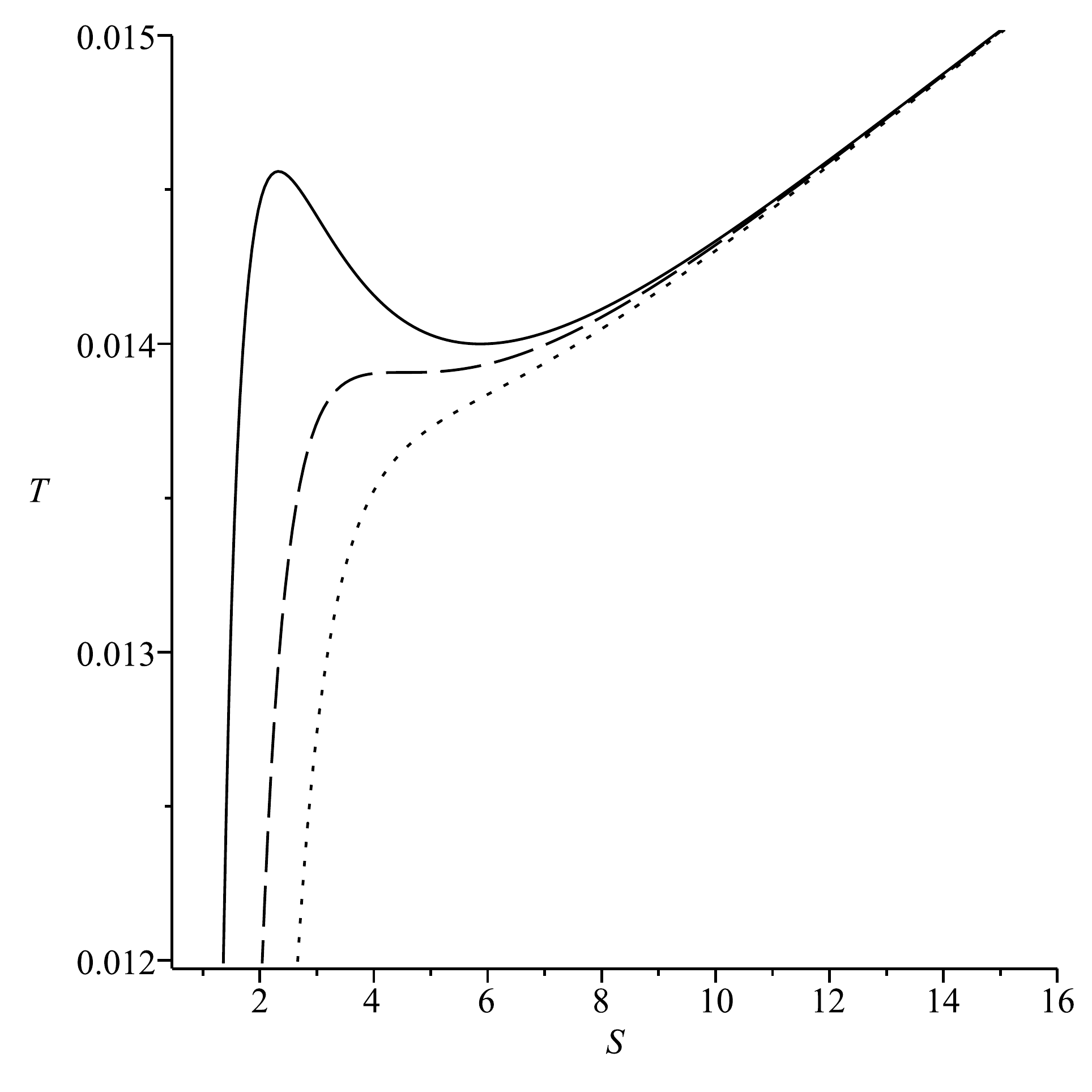}
\end{center}
\caption{Isocharge lines in the $T$-$S$ plane. For all the three plots, the parameters are fixed as $G=1,\beta=0.1,l=10$. Left plot: $n=2,d=4$;  middle plot: $n=3,d=6$; right plot: $n=4,d=8$. All the dashed lines   on the three plots correspond to $Q=Q_c$, all the solid lines correspond to $Q=0.6Q_c$, all the dotted lines correspond to $Q=1.4Q_c$.}
\label{fig3}
\end{figure}

\begin{figure}[h]
\begin{center}
\includegraphics[width=.30\textwidth]{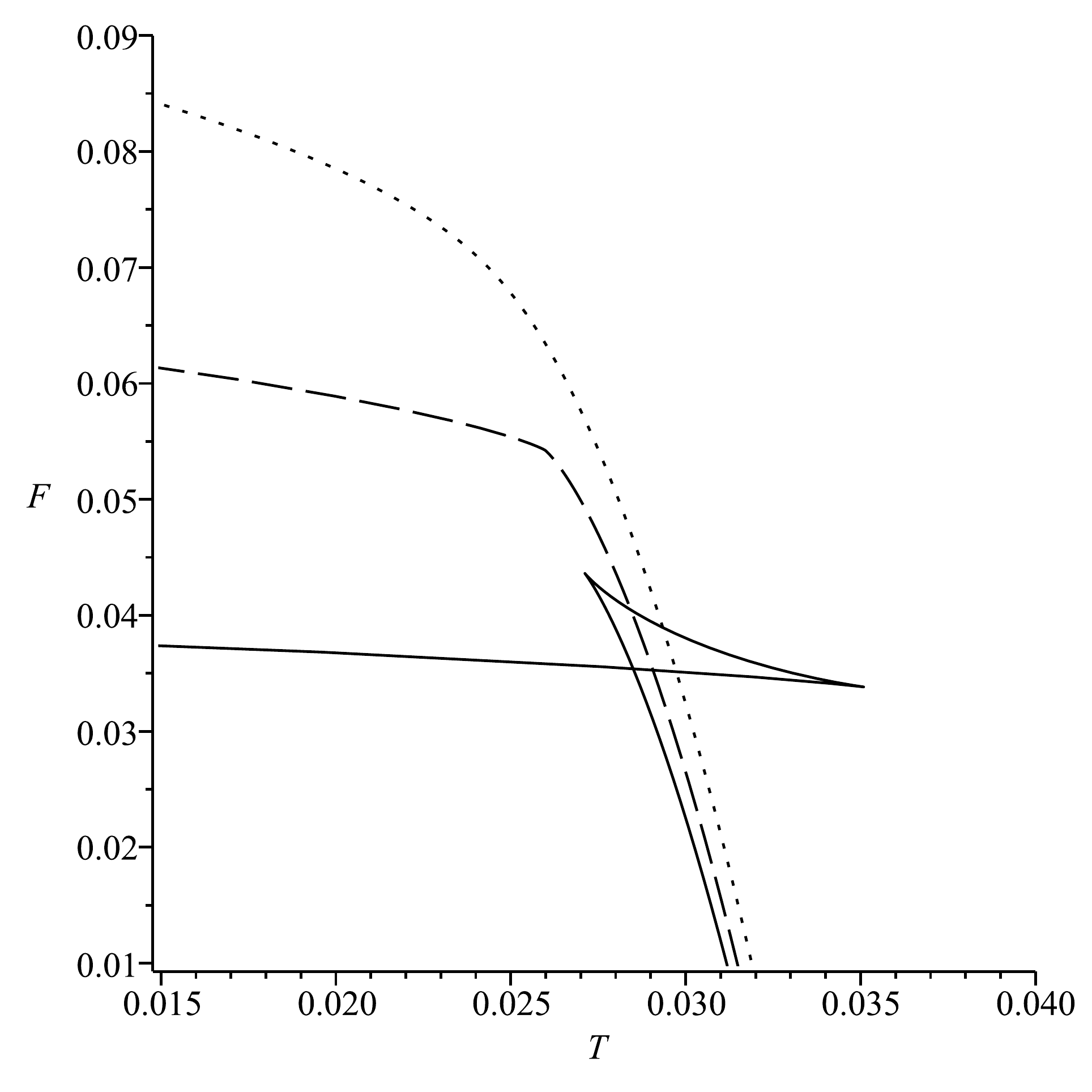}
\includegraphics[width=.30\textwidth]{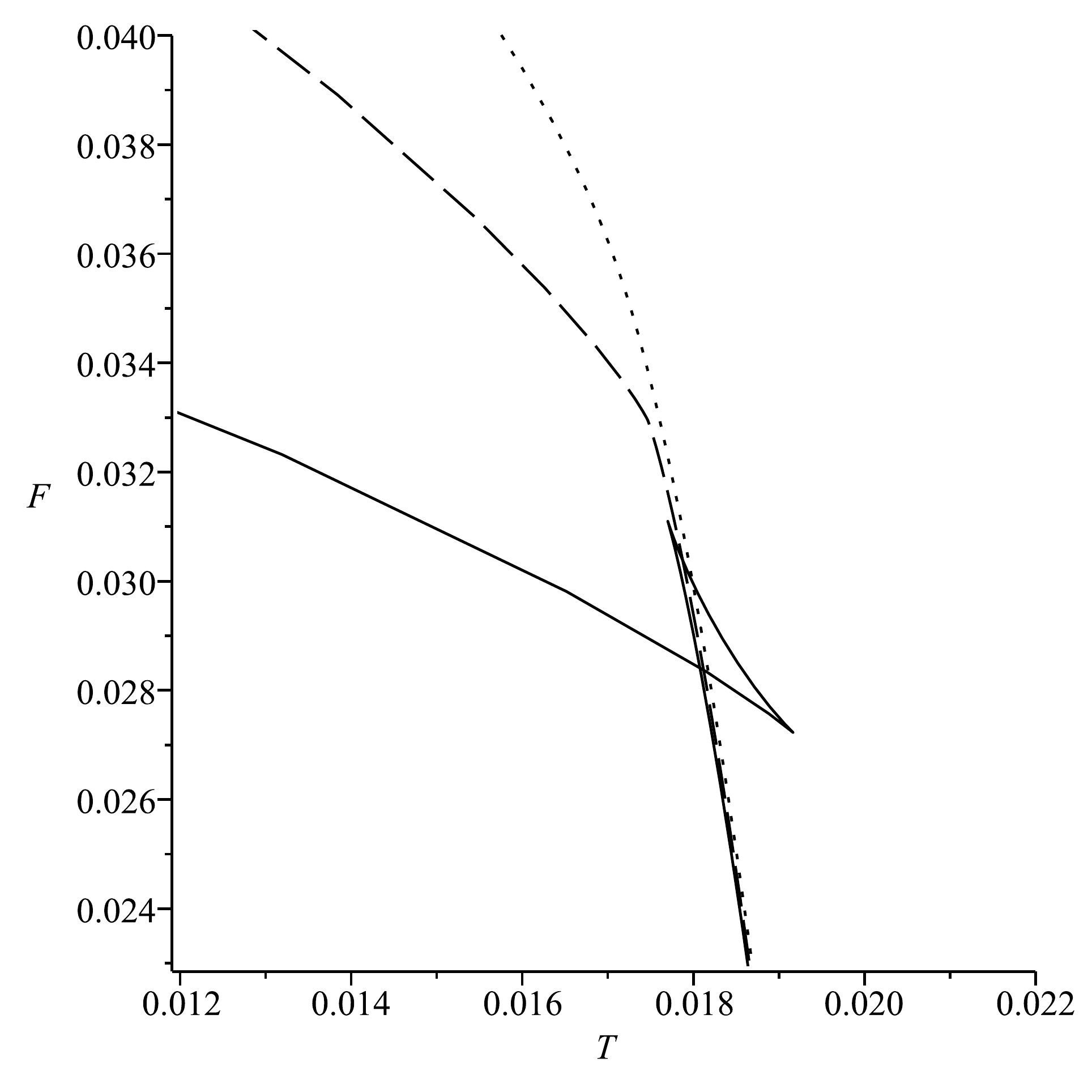}
\includegraphics[width=.30\textwidth]{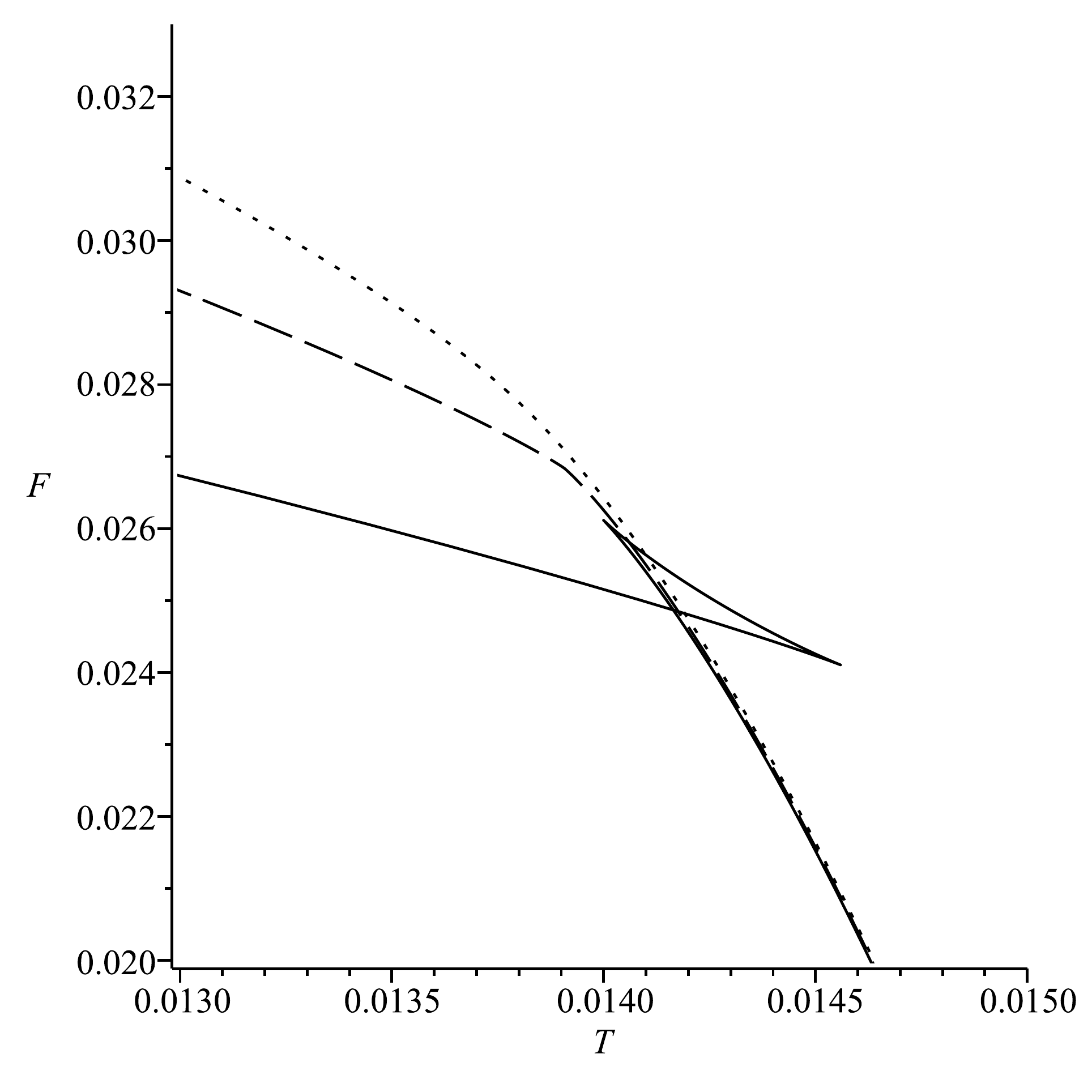}
\end{center}
\caption{Isocharge lines in the $F$-$T$ plane which are in one-to-one correspondence with the ones in the $T$-$S$ plane in Figure 3. Left plot: $n=2,d=4$; middle plot: $n=3,d=6$; right plot: $n=4,d=8$.}
\label{fig4}
\end{figure}

We can see from the table that, as the parameter $\beta$ increases the critical electric charge and the critical temperature decrease while the critical entropy increases, as the dimension of spacetime increases the critical electric charge and the critical entropy increase while the critical temperature decreases. The $T$-$S$ plots for $n=2,3,4$ are displayed in Fig.\ref{fig3}. We can also define the free energy
\begin{align}
F=M-TS,
\end{align}
and give the $F$-$T$ plots, as shown in Fig.\ref{fig4}. The solid, dashed and dotted lines in Fig.\ref{fig3} are in one-to-one correspondence to the solid, dashed and dotted ones in Fig.\ref{fig4}.

From Fig.\ref{fig3} we can see that, on each plot there exists a boundary curve (dashed isocharge line) corresponding to $Q=Q_c$ which describes an inflection point of a second order phase transition.
When $Q>Q_c$,  the isocharge line on each plot is always monotonous as shown by the dotted one, which implies the black hole is always stable, no phase transition occurs in this case. When $Q<Q_c$, as  the solid isocharge line  shows us, between two stable regions there is an unstable region where temperature decreases as entropy increases, i.e., heat capacity is negative, which implies phase transition occurs. The phase transition can be seen more manifestly on the $F$-$T$ plot, where we can see when $Q<Q_c$ ``swallow tail'' appears, for one temperature there are three black holes with different free energies, the black hole with the lowest free energies is the stable state. The unstable state transform to stable one via van der Waals-like  phase transition,  this kind of phase transitions is of first order.

For spatially flat black holes with $k=0$, substituting $Q^2$ obtained from $\frac{\partial T}{\partial S}\big|_Q=0$ into $\frac{\partial^2 T}{\partial S^2}\big|_Q$ we have
\begin{align}
\frac{\partial^2 T}{\partial S^2}\bigg|_Q&=C_{2}\cdot r_{+}^{-3},\;\;\;\;\;\;\;\;   n=2,\nonumber\\
\frac{\partial^2 T}{\partial S^2}\bigg|_Q&=C_3\cdot r_{+}^{-7}, \;\;\;\;\;\;\;\;  n=3,\label{criticaleqflat}\\
\frac{\partial^2 T}{\partial S^2}\bigg|_Q&=C_{4}\cdot r_{+}^{-11},  \;\;\;\;\;\;\; n=4,\nonumber
\end{align}
the specific form of $C_2, C_3, C_4$ are given in the appendix. The coefficients $C_2, C_3, C_4$ are constants independent of the horizon radius $r_{+}$, for higher dimensions the $C_n$'s are constants too. Thus $\frac{\partial^2 T}{\partial S^2}\big|_Q$ in (\ref{criticaleqflat}) do not vanish, which implies  there are no phase transitions for spatially flat even-dimensional black holes.

For hyperbolic black hole with $k=-1$,  numerical analysis indicates that if the critical equations are required to be satisfied, either the critical entropy or  square of the critical electric charge $Q_c^2$ is negative. Thus no phase transitions are found for even-dimensional hyperbolic black holes. This agrees with the result found in \cite{Kuang} for charged black holes of DCG with non-compact horizons.

\begin{figure}[h]
\begin{center}
\includegraphics[width=.30\textwidth]{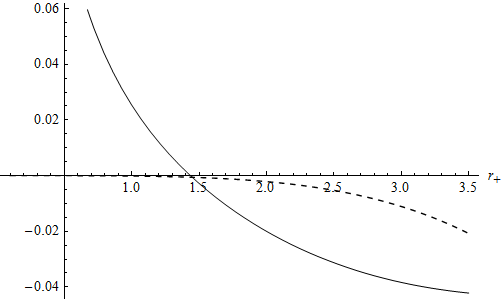}
\includegraphics[width=.30\textwidth]{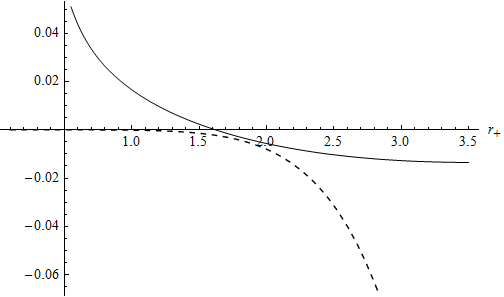}
\includegraphics[width=.30\textwidth]{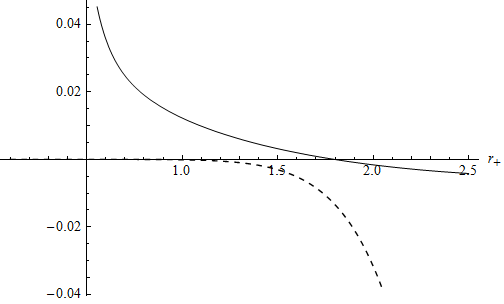}
\end{center}
\caption{The solid line on each plot describes $\frac{\partial^2T}{\partial S^2}\big|_Q$ versus $r_{+}$, the intersection of the solid line and the horizontal axis corresponds to the critical radius $r_c$. The dashed line on each plot describes $Q^2$ versus $r_{+}$, when $r_{+}$=$r_c$ then $Q=Q_c$. Left plot: $n=3$; middle plot: $n=4$; right plot:$n=5$.}
\label{fig5}
\end{figure}

Now let's discuss thermal phase transitions of odd-dimensional black holes.
When $k=+1$ and $n=2$,  from $\frac{\partial T}{\partial S}\big|_Q=0$ we get
\begin{align}
Q^2=-\frac{\beta r_{+}\sqrt{1+64\pi G l^2\beta^2}}{1+32\pi G l^2\beta^2}.
\end{align}
It can be seen that whatever value $r_{+}$ takes $Q^2$ is always negative, thus no phase transition occurs. For higher dimensions, numeral analysis indicates that, as displayed in Fig.\ref{fig5}, if the critical equations are required to be satisfied ($\frac{\partial T}{\partial S}\big|_{Q}=\frac{\partial^2 T}{\partial S^2}\big|_{Q}=0$), the square of critical electric charge $Q_c^2$ is negative. Thus no phase transition occurs in this case.
When $k=0$, $\frac{\partial^2 T}{\partial S^2}\big|_Q$ have similar form to Eq.(\ref{criticaleqflat}), thus there are no phase transitions in this case.
When $k=-1$, similar to the  even-dimensional hyperbolic black holes, either the critical entropy or the square of critical electric charge is negative,
no phase transition is found in this case either.
Therefore, no phase transitions are found for odd-dimensional black holes.

\section{Planar dyonic black hole\label{section3}}
\subsection{Local solution}
For the even dimensional spatially plat spacetime, we can also construct dyonic black hole solution. The BI Lagrangian density (\ref{LF1}) is not applicable for construction of dyonic black hole, we should adopt the following one\cite{1606.02733}
\begin{align}
L(F)=4\beta^2\sqrt{-\mathrm{det}(g_{\mu\nu})}-4\beta^2\sqrt{-\mathrm{det}\left(g_{\mu\nu}+\frac{F_{\mu\nu}}{\beta}\right)}.
\end{align}
EOM of the electromagnetic field now is
\begin{align}
\nabla_\mu\left[\frac{\sqrt{-h}}{\sqrt{-g}}\beta(h^{-1})^{[\mu\nu]}\right]=0,\label{EOMEM2}
\end{align}
and the energy momentum tensor is
\begin{align}
T^{\mu\nu}=2\beta^2 g^{\mu\nu}-2\beta^2\frac{\sqrt{-h}}{\sqrt{-g}} (h^{-1})^{(\mu\nu)}.\label{EMT2}
\end{align}
Where $h_{\mu\nu}\equiv g_{\mu\nu}+\frac{F_{\mu\nu}}{\beta}$, $h\equiv \mathrm{det} (h_{\mu\nu})$, $(h^{-1})^{\mu\nu}$ denotes the inverse of $h_{\mu\nu}$, satisfying
\begin{align}
(h^{-1})^{\mu\rho}h_{\rho\nu}=\delta^\mu_\nu,\;\;\;\;\;\;\;\;\;\;\;\;\;\;\;h_{\mu\rho}(h^{-1})^{\rho\nu}=\delta_\mu^\nu,
\end{align}
and
\begin{align}
(h^{-1})^{(\mu\nu)}=\frac{1}{2}\left[(h^{-1})^{\mu\nu}+(h^{-1})^{\nu\mu}\right],\;\;\;\;\;\;\;\;\;(h^{-1})^{[\mu\nu]}=\frac{1}{2}\left[(h^{-1})^{\mu\nu}-(h^{-1})^{\nu\mu}\right].
\end{align}

For $d=2+2j$ dimensional spacetime, we take the static metric ansatz
\begin{align}
ds^2=-f(r)dt^2+\frac{dr^2}{f(r)}+r^2(dx_1^2+dx_2^2+\cdots+dx_{2j-1}^2+dx_{2j}^2)\label{metricansatz2},
\end{align}
in order to construct dyonic black hole we take the field strength ansatz as
\begin{align}
F=-h'(r)dt\wedge dr+p(dx_1\wedge dx_2+\cdots+dx_{2j-1}\wedge dx_{2j}).\label{strength2}
\end{align}
Solving Eq.(\ref{EOMEM2}), we have
\begin{align}
h(r)=\int dr\frac{Q}{\sqrt{\frac{Q^2}{\beta^2}+\left(r^4+\frac{p^2}{\beta^2}\right)^n}}.\label{potential}
\end{align}
Substituting Eqs.(\ref{EMT2}-\ref{potential}) into Eq.(\ref{EOMG}), we obtain the dyonic black hole solution
\begin{align}
f(r)=\frac{r^2}{l^2}-r^2\left[\frac{16\pi GM}{\omega_2^jr^{d-1}}+64\pi G\beta r^{-1-2j}\int dr\left(-r^{2j}+\sqrt{Q^2+\left(r^4+p^2/\beta^2\right)^j\beta^2}\right)\right]^{\frac{1}{n-1}}.\label{dyonicbh}
\end{align}
where we define $\int dx_1dx_2=\cdots=\int dx_{2j-1}dx_{2j}=\omega_2$. Note that $d=2j+2=2n$. The explicit form of the solution (\ref{dyonicbh}) for $j=1$ is
\begin{align}
f(r)=&\frac{r^2}{l^2}-r^2\bigg[\frac{16\pi GM}{\omega_2^jr^{d-1}}+\frac{64}{3}\pi G \beta r^{-2} \bigg(-r^2+\sqrt{p^2+Q^2+r^4\beta^2}\nonumber\\
&+2\sqrt{p^2+Q^2}\;_2F_1\left[\frac{1}{4},\frac{1}{2},\frac{5}{4},-\frac{r^4\beta^2}{p^2+Q^2}\right]\bigg)\bigg],
\end{align}
and for $j=2$ is
\begin{align}
f(r)=&\frac{r^2}{l^2}-r^2\bigg[\frac{16\pi GM}{\omega_2^jr^{d-1}}-\frac{64}{5}\pi G r^{-4} \bigg(r^4\beta-5\sqrt{p^2+Q^2\beta^2}\nonumber\\
&\cdot F_1\bigg[\frac{1}{4},-\frac{1}{2},-\frac{1}{2},\frac{5}{4},-\frac{r^4\beta^2}{p^2+\sqrt{-Q^2\beta^2}},\frac{r^4\beta^2}{-p^2+\sqrt{-Q^2\beta^2}}\bigg]\bigg)\bigg]^{\frac{1}{2}},
\end{align}
where $F_1$ is the Appell hypergeometric function.

\subsection{Thermodynamics}
In this section, we calculate the thermodynamical quantities of the dyonic black hole and give the first law. We also study thermal phase transition of the black hole in $T$-$S$ plane.

In term of horizon radius, mass of the dyonic black hole is given by
\begin{align}
M=\frac{r_{+}^{2j+1}\omega_2^j}{16\pi G l^{2j}}-4\beta \omega_2^j \int_\infty^{r_{+}} dr\left(-r^{2j}+\sqrt{Q^2+\left(r^4+p^2/\beta^2\right)^j\beta^2}\right).
\end{align}
Note that we replace $d$ and $n$ with $j$ in the above equation. The temperature is
\begin{align}
T=\frac{f'(r_{+})}{4\pi}=\frac{r_{+}}{4j\pi l^2}\left(1+2j-64\pi G \beta l^{2j}r_{+}^{-2j}\left(-r_{+}^{2j}+\sqrt{Q^2+(r^4+p^2/\beta^2)^j\beta^2}\right)\right)
\end{align}
The Wald entropy of the black hole is
\begin{align}
S=\frac{l^{2-2j}r_{+}^{2j}\omega_2^j}{8G}.
\end{align}
The electric charge is calculated through performing integration of the flux of electromagnetic field in (\ref{EOMEM2}) on the $t=const$ and $r\rightarrow\infty$ hypersurface, yielding
\begin{align}
Q_e=4Q\omega_2^j.
\end{align}
The electric potential is given by
\begin{align}
\Phi_e=\int_{r_{+}}^\infty\frac{Q}{\sqrt{\frac{Q^2}{\beta^2}+\left(r^4+\frac{p^2}{\beta^2}\right)^j}}
\end{align}
The magnetic charge is
\begin{align}
Q_m=j\omega_2F_{x_1x_2}|_{r\rightarrow\infty}=j\omega_2p.
\end{align}
The first law is given by
\begin{align}
dM=TdS+\Phi_edQ_e+\Phi_mdQ_m,
\end{align}
where $\Phi_m$, which is defined as $\Phi_m=\frac{\partial M}{\partial Q_m}$, is the magnetic potential conjugate to the magnetic charge $Q_m$.

Now let's study phase transition of the dyonic black hole, we only consider the case $d=4$. By requiring the vanishing of
$\frac{\partial T}{\partial S}\big|_{Q,p}$ one obtains
\begin{align}
p^2=-Q^2+\frac{r_{+}^4\left[3(3+\Pi)+64\pi G\beta l^2\left(6+64\pi G l^2(\beta+2\beta^3)+\Pi\right)\right]}{8192 G^2 l^4 \pi^2\beta^2},\label{q2}
\end{align}
with
\begin{align}
\Pi=\sqrt{9+128\pi G\beta l^2\left(3+32\pi G \beta l^2(1+8\beta^2)\right)}.\label{Pi}
\end{align}
Substituting (\ref{q2}) into $\frac{\partial^2 T}{\partial S^2}\big|_{Q,p}$ we have
\begin{align}
\frac{\partial^2 T}{\partial S^2}\bigg|_{Q,p}=\frac{C}{G^{-2} \pi l^2\beta^3r_{+}^3\left[3(3+\Pi)+64\pi G\beta l^2(6+64\pi G l^2(\beta+4\beta^3)+\Pi)\right]^{3/2}\omega_2^2},\label{criticaldyonic}
\end{align}
where the numerator $C$ is a constant which is independent of $r_{+}$, the explicit form of $C$ is given in the Appendix. Thus, one can not solve out critical horizon by requiring
$\frac{\partial^2 T}{\partial S^2}\big|_{Q,p}=0$,  which implies there is no thermal phase transition for 4-dimensional(4d) dyonic black hole.

\section{Conclusions}
In this paper, we construct new topological black hole solutions of DCG coupled to BI electromagnetic field. In the limit $\beta\rightarrow\infty$ the black holes degenerate to charged black holes of DCG, in the limit $Q\rightarrow0$ the black holes degenerate to neutral black holes of DCG. In order to analyze the geometric aspect of the spacetime, we study the $r\rightarrow0$ asymptotic behaviors of $f(r)$ in detail, also we present the behaviors of $M(r_{+})$ versus $r_{+}$, the results obtained from these two methods mutually confirm. It is found that, if $d\neq 4h+2$, there exists a minimum $M_{ext}$ of $M$, at $r_{+}=0$, $M$ equals to some $M_0$. If $M<M_{ext}$, the spacetime describes a bare singularity, there is no horizon. If $M=M_{ext}$, there is a single horizon, this corresponds to an extremal black hole spacetime. If $M_{ext}<M<M_0$, there are two horizons. If $M\geq M_0$, there is one horizon. For $d=4h+2$, $M\geq M_0$ should be required in order to keep $f(r)$ being real, in this case $M(r_{+})$ is a monotonic function of $r_{+}$, one given mass corresponds to one horizon.

We calculate the conserved quantities of the topological black holes and check the first law is satisfied. We study thermal phase transitions of the black holes in $T$-$S$ plane.
We find for even-dimensional black holes with spherical topology, when $Q>Q_c$ the black holes are always stable. When $Q$ decreases to $Q_c$, the black holes undergo second order phase transitions. When $Q<Q_c$, the black holes undergo first order van der Waals-like phase transitions from unstable states to stable ones. For even-dimensional black holes with other topologies and odd-dimensional black holes with various topologies, no such phase transitions are found.

We also construct dyonic planar black holes in general even dimensions. We calculate the thermodynamical quantities of the dyonic black holes and give the first law. We study thermal phase transition of 4d dyonic black hole in $T$-$S$ plane, and find no phase transition occurs.

\section*{Acknowledgment}

This work is supported by the National Natural Science Foundation of China (NSFC) under the
grant numbers 11447153 and 11447196.

\section*{Appendix}
The coefficients in Eq.(\ref{criticaleqflat}) are given by
\begin{align}
C_2&=\frac{-96G^2}{\pi l^2  \Sigma_{d-2}^2 \left(-3-64\pi G l^2\beta^2+\sqrt{9+384\pi G l^2\beta^2(1+96\pi G l^2\beta^2)}\right)^3}\nonumber\\
&\cdot\bigg[\left(9+128\pi G l^2\beta^2(3+64\pi G l^2\beta^2)\right)\left(3+128\pi G l^2\beta^2(1+96\pi G l^2\beta^2)\right)\nonumber\\
&-\sqrt{3} (3+64\pi G l^2\beta^2)\left(3+128\pi G l^2\beta^2(1+64\pi G l^2\beta^2)\right)\sqrt{3+128\pi G l^2\beta^2(1+96\pi G l^2\beta^2)}\bigg]\nonumber\\
C_3&=\frac{8G^2l^2}{\pi  \Sigma_{d-2}^2 \left(5+64\pi G l^4\beta^2-\sqrt{25+128\pi G l^4\beta^2(5+1568\pi G l^4\beta^2)}\right)^3}\nonumber\\
&\cdot\bigg[\left(25+128\pi G l^4\beta^2(5+128\pi G l^4\beta^2)\right)\left(5+128\pi G l^4\beta^2(5+1568\pi G l^4\beta^2)\right)-\nonumber\\
&(5+64\pi G l^4\beta^2)\left(25+128\pi G l^4\beta^2(5+896\pi G l^4\beta^2)\right)\sqrt{25+128\pi G l^4\beta^2(5+1568\pi G l^4\beta^2)}\bigg]\nonumber,\\
C_4&=\frac{32G^2l^6}{9\pi  \Sigma_{d-2}^2 \left(7+64\pi G l^6\beta^2-\sqrt{49+128\pi G l^6\beta^2(7+3872\pi G l^6\beta^2)}\right)^3}\nonumber\\
&\cdot\bigg[\left(49+128\pi G l^6\beta^2(7+192\pi G l^6\beta^2)\right)\left(49+128\pi G l^6\beta^2(7+3872\pi G l^6\beta^2)\right)-\nonumber\\
&(7+64\pi G l^6\beta^2)\left(49+128\pi G l^6\beta^2(7+2112\pi G l^6\beta^2)\right)\sqrt{49+128\pi G l^6\beta^2(7+3872\pi G l^6\beta^2)}\bigg]\nonumber.
\end{align}
The numerator $C$ in Eq.(\ref{criticaldyonic}) is given by
\begin{align}
C=&-2\sqrt{2}(\sqrt{2}\big\{3+64\pi G\beta l^2\left[3(3+\Pi)+64\pi G\beta l^2(6+64\pi G l^2(\beta+4\beta^3)+\Pi)\right]^{3/2}\nonumber\\
&+2\big[81(3+\Pi)+64\pi G\beta l^2[81(4+\Pi)+64\pi G\beta l^2(4096\pi^2G^2\beta^2l^4(3+24\beta^2+16\beta^4)\nonumber\\
&+9(3+4\beta^2)(6+\Pi)+192\pi G\beta l^2(1+4\beta^2)(12+\Pi))]\big]\big\},\nonumber
\end{align}
with $\Pi$ given in Eq.(\ref{Pi}).

\providecommand{\href}[2]{#2}\begingroup
\footnotesize\itemsep=0pt
\providecommand{\eprint}[2][]{\href{http://arxiv.org/abs/#2}{arXiv:#2}}

\end{document}